\documentclass[journal,draftcls,onecolumn,12pt,twoside]{IEEEtran}
\usepackage{graphicx}
\usepackage{subfigure}
\usepackage{amsfonts, amssymb, amsthm}
\usepackage{amsmath}
\usepackage{enumerate}
\usepackage{multirow}
\usepackage{color}
\usepackage{cite}
\usepackage{diagbox}
\usepackage{xtab}
\theoremstyle{definition} \newtheorem{theorem}{Theorem}
\theoremstyle{definition} \newtheorem{corollary}[theorem]{Corollary}
\theoremstyle{definition} \newtheorem{proposition}[theorem]{Proposition}
\theoremstyle{definition} 
\theoremstyle{definition} \newtheorem{lemma}[theorem]{Lemma}
\theoremstyle{definition} 
\theoremstyle{definition} 
\theoremstyle{definition} \newtheorem{remark}{Remark}
\theoremstyle{definition} 
\theoremstyle{definition} \newtheorem*{example}{Example}
{\end{list}}

\begin{document}
\title{Delay-Complexity Trade-off of Random Linear Network Coding in Wireless Broadcast}
\author{\IEEEauthorblockN{Rina~Su\textsuperscript{\dag},~Qifu~Tyler~Sun\textsuperscript{\dag}{$^*$},~Zhongshan~Zhang\textsuperscript{\ddag}} \\
\IEEEauthorblockA{\textsuperscript{\dag}
Department of Communication Engineering, University of Science and Technology Beijing, P. R. China \\
\textsuperscript{\ddag} School of Information and Electronics, Beijing Institute of Technology, P. R. China \\
}
\thanks{$^*~$Q. T. Sun (Email: qfsun@ustb.edu.cn) is the corresponding author.}
}% <-this % stops a space
\maketitle
\thispagestyle{plain}
\pagestyle{plain}

\sloppy

\begin{abstract}

In wireless broadcast, random linear network coding (RLNC) over GF($2^L$) is known to asymptotically achieve the optimal completion delay with increasing $L$. However, the high decoding complexity hinders the potential applicability of RLNC schemes over large GF($2^L$). %
In this paper, a comprehensive analysis of completion delay and decoding complexity is conducted for field-based systematic RLNC schemes in wireless broadcast. In particular, we prove that the RLNC scheme over GF($2$) can also asymptotically approach the optimal completion delay per packet when the packet number goes to infinity. %
Moreover, we introduce a new method, based on circular-shift operations, to design RLNC schemes which avoid multiplications over large GF($2^L$). Based on both theoretical and numerical analyses, the new RLNC schemes turn out to have a much better trade-off between completion delay and decoding complexity. In particular, numerical results demonstrate that the proposed schemes can attain average completion delay just within $5\%$ higher than the optimal one, while the decoding complexity is only about $3$ times the one of the RLNC scheme over GF($2$).
\end{abstract}


\section{Introduction}
Wireless broadcast is an important transmission scenario with a variety of applications. For example, in a satellite or cellular network, a satellite or base station needs to broadcast a large file or streaming multimedia packets to many users in a timely manner. Such applications require cost efficient transmission techniques. Due to interference, fading or other channel imperfections, packet loss often happens and it results in receivers losing different subsets of packets during wireless broadcast. In the literature, linear network coding (LNC) has shown great ability to substantially improve transmission efficiency in wireless broadcast with erasures, and two main types of LNC techniques were extensively studied for wireless broadcast, namely random linear network coding (RLNC) \cite{Yu&Parastoo2018,Ioannis2017,Atilla2013,Emmanouil2016,NanXie2013,SkevakisL2016,SkevakisL2017,Lucani2012,Tsimbalo2018} %
and instantly decodable network coding (IDNC) \cite{Sadeghi2010,Aboutorab&Parastoo2014,Yu&Parastoo2014,Sameh2015,Wang2018}.

RLNC \cite{Ho2006} linearly combines all original packets at the sender with coefficients randomly and independently selected from a finite field to generate coded packets for transmission without feedback. IDNC \cite{Sadeghi2010, Sameh2010} can be regarded as a special type of deterministic LNC schemes, defined over GF($2$), for the wireless broadcast with feedback. Every coded packet generated by IDNC is specially designed based on receivers' side information obtained from the feedback, so as to enable as many receivers as possible to recover an original packet upon receiving the coded packet.

Though IDNC focuses more on the instantaneous decodability at receivers, compared with RLNC, it requires higher completion delay, which is a fundamental metric for transmission efficiency and refers to the total number of packets broadcast from the sender till all receivers successfully recover all original packets. %
It is well-known that when the finite field is sufficiently large, RLNC can achieve the optimal completion delay. However, a stringent issue that hinders the potential applicability of RLNC schemes over a large finite field is the very high decoding complexity. In the present paper, for wireless broadcast, we make a comprehensive analysis of completion delay and decoding complexity for field-based RLNC schemes and introduce a new method, based on circular-shift operations, to design RLNC schemes with a much better trade-off between completion delay and decoding complexity. Only systematic RLNC schemes are considered here because they can reduce the packet decoding delay and encoding/decoding complexity \cite{Lucani2012}\cite{Yu&Parastoo2014}\cite{Giuseppe&Tomaso2015}. The main contributions of the present paper are summarized as follows.
\begin{itemize}
\item For RLNC schemes over the finite field GF($2^L$), we provide explicit formulae for the expected completion delay and theoretically analyze the decoding complexity in terms of the total number of binary operations required for decoding. One may note that Ref. \cite{NanXie2013} analyzed the expected completion delay for a non-systematic RLNC scheme, but its analysis cannot directly imply the results herein since we focus on systematic RLNC schemes, which is more complicated to analyze than the non-systematic case.
\item It is well-known that RLNC schemes over GF($2^L$) can asymptotically achieve the optimal completion delay with increasing $L$. From another point of view, we prove that the RLNC scheme over GF($2$) can also asymptotically approach the optimal completion delay per packet with the increasing packet number. To the best of our knowledge, this intriguing property has never been revealed in the literature and cannot be deduced in a straightforward manner from other related works of RLNC for wireless broadcast.
\item Recent works \cite{Hou2016,Sun_TIT19,Sun_Commu19} advocated LNC design by adopting simple circular-shift operations in place of multiplications over a large finite field, and analyzed its performance in wireline networks which are totally different from wireless broadcast. %
    Inspired by them, we also propose a new method to design RLNC schemes based on circular-shifts. Different from them, the present new method does not introduce extra bits in the generation of coded packets, and the justification of the decodability of the proposed schemes without redundant bits is nontrivial. As a generalization of conventional RLNC schemes, the new method also introduces a new parameter $p_0$ to control the probability of coding coefficients to be zero.

\item We analyze our proposed schemes both theoretically and numerically, which verifies that the proposed schemes provide a much better trade-off between completion delay and decoding complexity. In particular, the expected completion delay of the proposed schemes is proved to be no larger than that of the conventional RLNC scheme over GF($q$) with $q \leq 1/p_0$. Moreover, numerical results demonstrate the average completion delay converges very fast to the optimal one (e.g., when the packet number $P \geq 15$, the average completion delay is just within $5\%$ higher than the optimal one), while the average number of binary operations for decoding is just about $3$ times that of the RLNC scheme over GF($2$).
\end{itemize}

The remainder of the paper is organized as follows. Section II first establishes the system model, and then analyzes the completion delay and decoding complexity of conventional field-based RLNC schemes. Section III introduces and analyzes our proposed RLNC schemes based on circular-shift operations. The performances are numerically compared in Section IV. Finally, Section V concludes the paper.

In addition to the proof details of some theorems and lemmas, the frequently used important notations are listed in Appendix D for reference.
%Due to space limit, all technical proofs herein are not provided. Please refer to the full version of the paper.
%Due to space limit, all technical proofs herein are not provided. Please refer to the full version of the paper in ArXiv.

\section{System Model And Analysis of Conventional RLNC Schemes}
\label{Sec:system_model}
\subsection{System Model}
Consider a single-hop erasure broadcast system without feedback, in which a unique sender  attempts to broadcast $P$ packets to $R$ receivers. Every packet consists of $M$ bits. In each timeslot, the sender can broadcast one packet to all receivers. The memoryless wireless channels between the sender and the receivers are subject to random packet erasures with erasure probability $1 - p_r$ at receiver $r$. Every receiver is interested in recovering all $P$ original packets.

All RLNC transmission schemes studied in this paper are systematic, that is, in the first phase of transmission, the sender will sequentially broadcast all $P$ original (uncoded) packets $\mathbf{m}_{1}, \mathbf{m}_{2}, \ldots, \mathbf{m}_P$. In the second phase of transmission, the sender will broadcast coded packets, which are linearly generated based on $P$ original packets, till every receiver can recover all $P$ original packets. The \emph{completion delay} $D$ of an RLNC scheme is the number of coded packets transmitted by the sender in phase two.

The conventional RLNC scheme is defined over the extension field GF($2^L$), where every packet is regarded as a row vector of $\frac{M}{L}$ symbols over GF($2^L$).\footnote{For simplicity, we assume $L$ divides $M$. In practice, $M$ is much larger than $L$ and dummy bits can be padded into a packet so that $L$ divides $M$.} Every packet broadcast by the sender is a GF($2^L$)-linear combination of $\mathbf{m}_{1}, \mathbf{m}_{2}, \ldots, \mathbf{m}_P$. Specifically, every coded packet $\mathbf{m}_{P+d}$, $d \geq 1$, randomly generated by the source in phase two can be expressed as
\begin{equation}
\mathbf{m}_{P+d} =\sum\nolimits_{j=1}^{P}\gamma_j \mathbf{m}_j,
\end{equation}
where the coding coefficients $\gamma_j$ are independently and uniformly selected from GF($2^L$). %
In order to indicate receivers how a packet $\mathbf{m}_{P+d} =\sum\nolimits_{j=1}^{P}\gamma_j \mathbf{m}_j$ is formed from the original $P$ packets, the \emph{coding vector} $\mathbf{f}_{P+d} = [\gamma_1, \ldots, \gamma_P]^{\mathrm{T}}$ consisting of $P$ coding coefficients is appended to $\mathbf{m}_{P+d}$ as a header. The coding vectors for $P$ original packets form an identity matrix of size $P$, \emph{i.e.}, $[\mathbf{f}_j]_{1 \leq j \leq P} = \mathbf{I}_P$. When a receiver $r$ obtains $P$ packets whose coding vectors are linearly independent, the $P$ original packets can be decoded at $r$.

When $L$ is large enough, the RLNC scheme over GF($2^L$) can be approximately regarded as the \emph{perfect} RLNC scheme \cite{Sameh2015}, in which arbitrary $P$ packets generated by the source have linearly independent coding vectors. The prefect RLNC scheme is optimal in terms of completion delay $D$. However, the computational complexity of this scheme in both encoding and decoding is high because multiplications in GF($2^L$) have to be invoked. %
On the contrary, the RLNC scheme over GF($2$), \emph{i.e.} with $L = 1$, only requires bitwise additions among packets in both encoding and decoding. As a result, it involves lowest computational complexity among RLNC schemes, but its completion delay $D$ is relatively higher.

As the conventional RLNC schemes will be the benchmarks for our new proposed RLNC schemes, the following two subsections theoretically analyze their performance in terms of completion delay and decoding complexity.

\subsection{Completion Delay Analysis}
Throughout the paper, unless otherwise specified, for a considered RLNC scheme, let $D_r$ denote the number of coded packets transmitted in phase two by the source till receiver $r$ is able to successfully recover all $P$ original packets. Then, the completion delay $D$ can be defined as
\begin{equation}
D = \max\{D_1, D_2, \ldots, D_R\},
\end{equation}
and the expectation of $D$ can be characterized as
\begin{equation}
\label{eqn:expection_D_general_formula}
  \mathbb{E}[D] = \sum _{d\geq 0}\mathrm{Pr}(D > d)=\sum_{d\geq 0}\left(1-\prod_{1\leq r \leq R }\mathrm{Pr}(D_r \leq d)\right)
\end{equation}

Consider the conventional RLNC scheme over GF($q$), where $q = 2^L$. %
At receiver $r$, denote by $U_{r}$ the number of successfully received uncoded packets in phase one, and assume that for the $U_r$ uncoded packets and all the coded packets successfully received, the columnwise juxtaposition of their coding vectors forms a matrix of rank $U_r + j$. %
When receiver $r$ successfully obtains a new coded packet, note that the probability for the new packet to be \emph{innovative}, \emph{i.e.}, its coding vector is linearly independent of the previous ones, is $1-q^{U_r+j-P}$. %
For brevity, write
\begin{equation}
\label{eqn:p_prim_definition}
p'_{r,U_r+j} = p_{r}(1-q^{U_r+j-P}).
\end{equation}

Assume $U_r = u$. When $u = P$, all the original packets have already been successfully received during the first $P$ rounds of transmissions, so that $D_r = 0$. When $u < P$, receiver $r$ still requires to receive $P-u$ innovative coded packets. Thus, $D_r$ can be expressed as
\begin{equation}
\label{eqn:D_r_expression}
D_r = A_1 + A_2 +\ldots + A_{P-u},
\end{equation}
where $A_j$, $1 \leq j \leq P-u$, represents the number of packets obtained by receiver $r$ during the process that the number of innovative packets at receiver $r$ increases from $u+j-1$ to $u+j$. Thus, for $a_j \geq 1$,
\begin{equation}
\mathrm{Pr}(A_j = a_j) = (1 - p'_{r, u+j-1})^{a_j-1}p'_{r, u+j-1}.
\end{equation}
Consequently, for $d > 0$, the conditional probability of $D_r$ is equal to $d$ given $U_r = u$ can be characterized as
\begin{align}
\mathrm{Pr}(D_r = d | U_r = u) = \left\{\begin{matrix}
0, & u = P, u < P-d \\
\label{eqn:conditional_pmf_RLNC_GF2}
\sum\limits_{\mathbf{a} \in \mathcal{A}_{P-u,d}} \prod\limits_{j=1}^{P-u}(1 - p'_{r, u+j-1})^{a_j-1}p'_{r, u+j-1}, & \mathrm{otherwise}
\end{matrix}\right.
\end{align}
where $\mathcal{A}_{P-u,d}$ is defined as the collection of $(P-u)$-tuples $\mathbf{a}=(a_{1},\ldots, a_{P-u})$ subject to $a_{1}, \ldots, a_{P-u} \in \mathbb{Z}^{+}$ and $a_{1} + \ldots + a_{P-u} = d$. Note that $u < P-d$ expresses the case that the number of successfully received packets by receiver $r$ is less than $P$, not sufficient to recover all $P$ original packets. %
Based on (\ref{eqn:conditional_pmf_RLNC_GF2}), the probability distribution of $D_{r}$ can be formulated as $\mathrm{Pr}(D_r = 0) = p_r^P$ and for $d > 0$,

%\begin{align}
%&\mathrm{Pr}(D_r = d) = \nonumber \\
%&\sum_{u = \max\{0, P-d\}}^{P-1} \binom{P}{u} p_r^u(1-p_r)^{P-u}\mathrm{Pr}(D_r = d | U_r = u).
%\label{eqn:pmf_RLNC_GF2}
%\end{align}
\begin{equation}
\mathrm{Pr}(D_r = d) =
\sum_{u = \max\{0, P-d\}}^{P-1} \binom{P}{u} p_r^u(1-p_r)^{P-u}\mathrm{Pr}(D_r = d | U_r = u).
\label{eqn:pmf_RLNC_GF2}
\end{equation}

By plugging (\ref{eqn:pmf_RLNC_GF2}) into (\ref{eqn:expection_D_general_formula}), we can get a formula to compute the expected value of $D$ for the conventional RLNC scheme over GF($q$). Even though the computation of $\mathbb{E}[D]$ in terms of (\ref{eqn:expection_D_general_formula}) involves an infinite term summation, the term $1 - \prod_{1 \leq r \leq R} \mathrm{Pr}(D_r \leq d)$ converges to $0$ quickly when $d$ increases. Therefore, a very close approximate for $\mathbb{E}[D]$ can be obtained based on (\ref{eqn:expection_D_general_formula}) when $P$ is not large.

Note that the above analysis on $\mathbb{E}[D]$ also applies to an odd prime power $q$, though such $q$ is not the focus of this paper.

When $q$ increases, $p'_{r,U_r+j}$ tends to $1$ for all $U_r+j < P$, so that the completion delay of the RLNC scheme over GF($q$) approaches to the one of the perfect RLNC scheme, where as long as $P$ packets are received by receiver $r$, the original $P$ packets can be recovered. %
For the perfect RLNC scheme, $P+D_r$ follows the negative binomial distribution with the probability mass function as
$\mathrm{Pr}(D_r = d) = \binom{P+d-1}{P-1} p_r^P(1-p_r)^d, d \geq 0$. %
Recall that for positive integers $a$ and $b$, the \emph{regularized incomplete beta function} $I_x(a,b+1)$ can be expressed in the following combinatorial manner $I_x(a,b+1) = \sum_{j=0}^{b} \binom{a+j-1}{a-1} x^a(1-x)^{j}$. %
As a result, for the perfect RLNC scheme, $\mathrm{Pr}(D_r \leq d) = I_{p_r}(P, d+1)$, and
\begin{equation}
\mathbb{E}[D] = \sum\nolimits_{d \geq 0} (1 - \prod\nolimits_{1 \leq r \leq R} I_{p_r}(P, d+1) ).
\end{equation}

For a given packet number $P$, let $D^{\mathrm{GF}(2^L)}$,  $D^{\mathrm{perf}}$ respectively denote the completion delay of the RLNC scheme over GF($2^L$) and the perfect RLNC scheme. It is well known that $D^{\mathrm{GF}(2)}$ has the largest value. %
However, the next theorem unveils an inherent connection between $D^{\mathrm{GF}(2)}$ and $D^{\mathrm{perf}}$.

\begin{theorem}
\label{thm:GF2_scheme_convergence}
$\lim_{P\rightarrow\infty}\mathbb{E}[D^{\mathrm{GF}(2)}]/P = \lim_{P\rightarrow\infty}\mathbb{E}[D^{\mathrm{perf}}]/P$.
\begin{IEEEproof}
Please refer to Appendix-\ref{appendix:proof_thm:GF2_scheme_convergence}.
\end{IEEEproof}
\end{theorem}

\begin{remark}
Same as in the proof of Theorem \ref{thm:GF2_scheme_convergence}, let $N_r$ denote the number of successfully received packets at a single receiver $r$ till $r$ can recover all $P$ original packets. For non-systematic RLNC, an upper bound on $\mathbb{E}[N_r]$ was derived in \cite{Lucani2012} and was recently tightened in \cite{Ioannis2017}. %
In particular, regardless of the choice of GF($2^L$), $\mathbb{E}[N_r] \leq P + 2$. Based on this, it is straightforward to deduce that $\lim_{P\rightarrow\infty}\mathbb{E}[D_r^{\mathrm{GF}(2)}]/P = \lim_{P\rightarrow\infty}\mathbb{E}[D_r^{\mathrm{perf}}]/P$. %
However, it is insufficient to further imply Theorem \ref{thm:GF2_scheme_convergence}, which considers the completion delay $D$ of the whole system instead of a single receiver. To the best of our knowledge, there was no claim similar to Theorem \ref{thm:GF2_scheme_convergence} in the literature.
\end{remark}

Theorem \ref{thm:GF2_scheme_convergence} is of theoretical interest because it implies that for a fixed and extremely large packet number, though impractical, the RLNC scheme over GF($2$) is optimal for both completion delay per packet and decoding complexity.

\subsection{Decoding Complexity Analysis}
For the conventional RLNC scheme over GF($2^L$), assume that receiver $r$ has successfully obtained $U_r$ uncoded packets $\mathbf{m}_1', \ldots, \mathbf{m}_{U_r}'$ and $P- U_r$ coded packets $\mathbf{m}_{U_r+1}', \ldots, \mathbf{m}_{P}'$ whose coding vectors are linearly independent. We next analyze the number of binary operations required to recover the original $P$ packets $\mathbf{m}_1, \ldots, \mathbf{m}_P$ from $\mathbf{m}_1', \ldots, \mathbf{m}_P'$.
 Assume $\mathbf{m}_j = \mathbf{m}_j'$ for all $1 \leq j \leq U_r$. %

Further, in this subsection, denote the coding vector for $\mathbf{m}'_j$ by $[\gamma_{j1} \ldots \gamma_{jP}]^{\mathrm{T}}$, so that $\mathbf{m}'_j = \sum_{1 \leq j' \leq P} \gamma_{jj'}\mathbf{m}_{j'}$.

The decoding process consists of two phases. In phase \uppercase\expandafter{\romannumeral1}, receiver $r$ removes the information of uncoded packets $\mathbf{m}_1', \ldots, \mathbf{m}_{U_r}'$ from coded packets $\mathbf{m}_{U_r+1}', \ldots, \mathbf{m}_{P}'$. Specifically, for $U_r < j \leq P$, reset $\mathbf{m}_{j}'$ as
\begin{equation}
\label{eqn: procedure phase I}
\mathbf{m}_j' = \mathbf{m}_j' - \sum\nolimits_{1 \leq j' \leq U_r} \gamma_{jj'}\mathbf{m}_{j'}'.
\end{equation}
After phase I, the coded packets $\mathbf{m}_{U_r+1}', \ldots, \mathbf{m}_P'$ are purely linear combinations of $\mathbf{m}_{U_r+1}, \ldots, \mathbf{m}_P$. Write $\mathcal{A} = \mathcal{A}' = \{U_r+1, \ldots, P\}$.

The decoding process of phase \uppercase\expandafter{\romannumeral2} consists of two steps to recover $\mathbf{m}_{U_r+1}, \ldots, \mathbf{m}_P$ from $\mathbf{m}_{U_r+1}', \ldots, \mathbf{m}_P'$. %
In the first step of phase II, recursively perform the following procedure. From $\mathcal{A}$, select an index, say $j_0$, such that there is only one nonzero coefficient, say $\gamma_{j_0j_0'}$, in $\{\gamma_{j_0j'}: j' \in \mathcal{A}'\}$. Thus, $\mathbf{m}_{j_0}' = \gamma_{j_0j_0'}\mathbf{m}_{j_0'}$. %
Remove $j_0$ and $j_0'$ from $\mathcal{A}$ and $\mathcal{A}'$ respectively. Recover $\mathbf{m}_{j_0'}$ by the operation $\gamma_{j_0j_0'}^{-1}\mathbf{m}_{j_0}'$. For every remaining coded packet $\mathbf{m}_{j}'$, $j \in \mathcal{A}$, reset it by subtracting $\mathbf{m}_{j_0'}$ from $\mathbf{m}_{j}'$ via
\begin{equation}
\label{eqn: procedure phase II A}
\mathbf{m}_{j}' = \mathbf{m}_{j}' - \gamma_{jj_0'}\mathbf{m}_{j_0'},
\end{equation}
so that $\mathbf{m}_j'$ keeps equal to $\sum_{j'\in \mathcal{A}'}\gamma_{jj'}\mathbf{m}_{j'}$ for all $j \in \mathcal{A}$. %
The above recursive procedure ends till for every $j \in \mathcal{A}$, there are at least two nonzero coefficients in $\{\mathbf{\gamma}_{jj'}: j' \in \mathcal{A}'\}$.

After the first step of phase II decoding, if $\mathcal{A}$ is empty, it means all $P$ original packets have been successfully recovered. Otherwise, continue to perform the second step. In step two, first compute the inverse matrix of the square matrix $[\gamma_{jj'}]_{j\in \mathcal{A}, j'\in\mathcal{A}'}$\footnote{Throughout the paper, $[\mathbf{A}_{ab}]_{a,b}$ denotes the matrix generated by $\mathbf{A}_{ab}$, with $a$ and $b$ respectively representing the column and the row index.} of size $|\mathcal{A}|$. Let $\mathbf{D} = [\beta_{j'j}]_{j'\in \mathcal{A}',j \in \mathcal{A}}$ denote this inverse matrix. Then, the original packet $\mathbf{m}_{j'}$, $j' \in \mathcal{A}'$, can be recovered by
\begin{equation}
\label{eqn: procedure phase II B}
\mathbf{m}_{j'} = \sum\nolimits_{j \in \mathcal{A}} \beta_{j'j}\mathbf{m}_{j}'.
\end{equation}

We now analyze the binary operation number involved in the above decoding process. Follow the same consideration in \cite{Sun_TIT19}, assume it takes $L$ binary operations to compute the addition of two elements in GF($2^L$), and at least $2L^2$ binary operations to compute the multiplication of two elements in GF($2^L$). %
Since every packet contains $\frac{M}{L}$ symbols over GF($2^L$), it takes at least $\frac{M}{L}(2L^2+L)$ binary operations to compute one term $\mathbf{m}_j' - \gamma_{jj'}\mathbf{m}_{j'}'$ for nonzero $\gamma_{jj'}$ in (\ref{eqn: procedure phase I}). In addition, since the probability for $\gamma_{jj'}$ to be nonzero equals to $(2^L-1)/2^L$, it can be easily deduced that the number of binary operations required for receiver $r$ in the decoding process of phase I is
\begin{align}
\label{general eqn: complexity of phase I}
&U_r(P-U_r)\frac{2^L-1}{2^L}\frac{M}{L}(2L^2+L)
% =& MU_r(P-U_r)(2L+1)\frac{2^L-1}{2^L}.
\end{align}

In the first step of phase II decoding, the recursive procedure takes $P-U_r-|\mathcal{A}|$ iterations, where $\mathcal{A}$ refers to the index set defined at the end of the first step. The number of binary operations required in this step at receiver $r$ is
\begin{equation}
\label{general complete eqn: complexity of phase II A}
\sum\nolimits_{j=|\mathcal{A}|+1}^{P-U_r} (\frac{M}{L}(2L^2)+(j-1)\frac{2^L-1}{2^L}\frac{M}{L}(2L^2+L))
\end{equation}
In the second step of phase II decoding, it takes additional
\begin{equation}
\label{general eqn: complexity of phase II B}
\phi(\mathbf{D})\frac{M}{L}(2L^2)+(\phi(\mathbf{D}) - |\mathcal{A}|)\frac{M}{L}L
\end{equation}
binary operations to recover the last $|\mathcal{A}|$ original packets according to (\ref{eqn: procedure phase II B}), where $\phi(\mathbf{D})$ refers to the number of nonzero entries in the decoding matrix $\mathbf{D}$. %
We remark here that the computational complexity to compute the inverse matrix $\mathbf{D}$ is ignored, because in practice the packet length $M$ and the number of symbols $\frac{M}{L}$ is much larger than the size $|\mathcal{A}|$ of $\mathbf{D}$.

For the case that $L$ is large, since every random coding coefficient is nonzero with high probability, the probability to perform step one of phase II decoding is low. As a consequence, we can assume that the decoding matrix $\mathbf{D}$ is of size $P-U_r$, and $\phi(\mathbf{D}) = (P-U_r)^2$. Based on (\ref{general eqn: complexity of phase I}) and (\ref{general eqn: complexity of phase II B}), the number of binary operations for decoding is approximated as $M[(2L+1)P-1](P-U_r)$. As $\mathbb{E}[U_r]=Pp_r$ and $\mathbb{E}[U_r^2] = Pp_r(Pp_r - p_r +1)$, the expected number of binary operations for decoding is
\begin{equation}
MP[(2L+1)P-1](1-p_r)
\end{equation}

For the case $L = 1$, as there is no need to do multiplication in an extension field, it takes $U_r M$ binary operations to compute (\ref{eqn: procedure phase I}). In addition, instead of (\ref{general eqn: complexity of phase II B}), where $\phi(\mathbf{D})$ needs to be approximated, we can choose an alternative lower bound $M(|\mathcal{A}|-1)$ on the number of binary operations required to recover the last $|\mathcal{A}|$ original packets from the coded packets in $\mathcal{A}$. In total, the expected number of binary operations for decoding is lower bounded by
\begin{equation}
\label{eqn:decoding_complexity_GF2}
\frac{M}{2}\left[(P^2-P)p_r(1-p_r)+
3|\mathcal{A}|-2\right].
\end{equation}

\section{Proposed Circular-Shift RLNC Schemes}
For RLNC schemes over large GF($2^L$), though it can approach a near optimal completion delay, the decoding complexity becomes much higher because the decoding process involves heavy multiplications in GF($2^L$). One possible way to alleviate the decoding complexity imposed by extension field multiplication is to use sparse encoding vectors. This inspired us to increase the probability of zero coding coefficients in RLNC scheme design. Furthermore, in order to avoid extension-field multiplication, Ref. \cite{Hou2016,Sun_TIT19,Sun_Commu19} motivate us to adopt circular-shift operations in RLNC design. It turns out that compared with the perfect RLNC scheme, the
proposed \emph{circular-shift RLNC} schemes, have comparative completion delay but a much lower decoding complexity.

\subsection{Scheme Description}
Hereafter in this paper, let $L$ be an even integer such that $L+1$ is a prime with primitive root $2$, that is, the multiplicative order of $2$ modulo $L+1$ is equal to $L$. Denote by $\mathbf{C}_{L+1}$ the $(L+1)\times (L+1)$ cyclic permutation matrix $\begin{bmatrix} \mathbf{0} & \mathbf{I}_{L} \\ 1 & \mathbf{0} \end{bmatrix}$. %
For a binary sequence $\mathbf{m}$ of length $L+1$, the linear operation $\mathbf{m}\mathbf{C}_{L+1}^l$, $1 \leq l \leq {L+1}$, is equivalent to a circular-shift of $\mathbf{m}$ by $l$ bits to the right.  % \emph{i.e.},
%\begin{align}
%\label{eqn: linear operation of C_L}
%[m_1,\ldots, m_{L+1}]\mathbf{C}_{L+1}^l=[m_{L-l+2},\ldots,m_{L+1}, m_1,\ldots, m_{L-l+1}]
%\end{align}
Define another $L \times (L+1)$ matrix $\mathbf{G}$ and $(L+1) \times L$ matrix $\mathbf{H}$ over GF($2$) by
\begin{equation}
\label{eq:definition_GH}
\mathbf{G} = [\mathbf{I}_L~\mathbf{1}], \mathbf{H} = [\mathbf{I}_L~\mathbf{0}]^{\mathrm{T}},
\end{equation}
\emph{i.e.}, the first $L$ columns in $\mathbf{G}$ and the first $L$ rows in $\mathbf{H}$ are identity matrix, while all entries in the last column in $\mathbf{G}$ equal to $1$ and all entries in the last row in $\mathbf{H}$ equal to $0$. %
\footnote{Throughout the paper, $\mathbf{1}$ and $\mathbf{0}$ refer to an all-one matrix and an all-zero matrix, respectively. The size of $\mathbf{0}$ or $\mathbf{1}$, if not explicitly explained, can be inferred in the context.}
Write
\begin{equation}
\label{eq:definition_C_cal}
\mathcal{C} = \{\mathbf{0}, \mathbf{G}\mathbf{C}_{L+1}\mathbf{H}, \mathbf{G}\mathbf{C}_{L+1}^2\mathbf{H}, \ldots, \mathbf{G}\mathbf{C}_{L+1}^{L+1}\mathbf{H} \}.
\end{equation}
Note that $\mathbf{G}\mathbf{C}_{L+1}^{L+1}\mathbf{H} = \mathbf{I}_{L}$.

Same as in the previous section, regard every packet $\mathbf{m}_j$ of $M$ bits as a row vector $[\mathbf{s}_{j,1}, \mathbf{s}_{j,2}, \ldots, \mathbf{s}_{j,\frac{M}{L}}]$ of $\frac{M}{L}$ symbols, where every symbol $\mathbf{s}_{j,j'}$ itself is merely regarded as an $L$-dimensional row vector over GF($2$) rather than an element in GF($2^L$). For an $L \times L'$ matrix $\mathbf{A}$, denote by $\mathbf{A} \circ \mathbf{m}_j$ the following linear operation on $\mathbf{m}_j$
\begin{equation}
\label{eqn:circ-symbol-definition}
\mathbf{A} \circ \mathbf{m}_j = [\mathbf{s}_{j,1}\mathbf{A}, \mathbf{s}_{j,2}\mathbf{A}, \ldots, \mathbf{s}_{j,\frac{M}{L}}\mathbf{A}],
\end{equation}
that is, to perform symbol-wise multiplication by coefficient $\mathbf{A}$ on $\mathbf{m}_j$, so that each of the $\frac{M}{L}$ new symbols has length $L'$.

Same as in the conventional systematic RLNC schemes, for the proposed circular-shift RLNC scheme, the sender will first sequentially broadcast $P$ original $M$-bit packets $\mathbf{m}_{1}, \mathbf{m}_{2}, \ldots, \mathbf{m}_P$ and then random linear combinations $\mathbf{m}_{P+1}, \mathbf{m}_{P+2}, \ldots, \mathbf{m}_{P+D}$ of the $P$ original packets till every receivers can recover all $P$ original packets. The random linear combinations are based on the linear operations $\circ$ with coefficients selected from $\mathcal{C}$. Specifically, every coded packet $\mathbf{m}_{P+d}$, $d \geq 1$, can be expressed as
\begin{equation}
\label{eqn:C-S RLNC encoding}
\mathbf{m}_{P+d} =\sum\nolimits_{j=1}^{P}\mathbf{\Gamma}_j \circ \mathbf{m}_j,
\end{equation}
where the coding coefficients $\mathbf{\Gamma}_j$ are randomly and independently selected from $\mathcal{C}$ according to
\begin{align}
\label{eqn: probability distribution of coding coefficents}
\mathrm{Pr}(\mathbf{\Gamma}_j=\mathbf{\Gamma})=
&\left\{\begin{matrix}
p_0, & \mathrm{\mathbf{\Gamma} = \mathbf{0}} \\
\frac{1-p_0}{L+1}, & \mathbf{\Gamma}\in \mathcal{C}\backslash\{\mathbf{0}\}
\end{matrix}\right.
\end{align}
Note that when coding coefficients are generated, there is a particular parameter $p_0$ to control the probability of $0$ to occur. The more frequent occurrence of $0$ in the coding coefficients will help reduce the decoding complexity at receivers, but will increase the completion delay. We shall assume that $p_0$ is a rational number no smaller than $1/(L+2)$. When $p_0 = 1/(L+2)$, the probability to choose $\mathbf{0}$ is the same as to choose a particular nonzero coding coefficient from $\mathcal{C}$. %

Since the proposed RLNC schemes inherit the linearity among vectors with the (matrix) coefficients selected from $\mathcal{C}$, they belong to \emph{vector linear network coding} \cite{Sun_Commu16}.

By a slight abuse of notation, define the \emph{coding vector} of a packet to be a $PL\times L$ matrix over GF($2$) as follows. Every coding vector can be regarded as a $P \times 1$ block matrix which is a coding coefficient (matrix) belonging to $\mathcal{C}$. For an original (uncoded) packet $\mathbf{m}_j$, $1 \leq j \leq P$, the $j^{\mathrm{th}}$ block in its coding vector $\mathbf{F}_j$ is the identity matrix $\mathbf{I}_L$, while all other blocks are $L\times L$ zero matrices. Thus, $[\mathbf{F}_j]_{1\leq j \leq P} = \mathbf{I}_{PL}$. %
For a coded packet $\mathbf{m}_{P+d} =\sum\nolimits_{j=1}^{P}\mathbf{\Gamma}_j \circ \mathbf{m}_j$, its coding vector $\mathbf{F}_{P+d}$ is defined as $\mathbf{F}_{P+d} = [\mathbf{\Gamma}_1^{\mathrm{T}}~\mathbf{\Gamma}_2^{\mathrm{T}}~\ldots~\mathbf{\Gamma}_P^{\mathrm{T}}]^{\mathrm{T}}$. As there are only $L+2$ possible choices for coding coefficients, only $\lceil P\log_2(L+2)\rceil$ bits are sufficient in the packet header to store the coding coefficient information.

For $J \geq 2$, assume that receiver $r$ has successfully received $J-1$ packets, and denote by $\mathbf{F}$ the $PL\times (J-1)L$ matrix over GF($2$) obtained by column-wise juxtaposition of the coding vectors for the $J-1$ packets. For the $J^{\mathrm{th}}$ successfully received packet with the coding vector $\mathbf{F}_{J}$, it is said to be \emph{innovative} if $\mathrm{rank}([\mathbf{F}~\mathbf{F}_{J}]) - \mathrm{rank}(\mathbf{F}) = L$. As long as a receiver obtains $P$ innovative packets, that is, their coding vectors can be columnwise juxtaposed into a $PL\times PL$ full rank matrix over GF($2$), the $P$ original packets can be decoded at $r$. %

We next analyze the completion delay of the proposed circular-shift RLNC scheme.  %
 Note that even though the works in \cite{Hou2016,Sun_TIT19,Sun_Commu19} studied different LNC schemes based on circular-shift operations, they targeted at wireline network models, totally different from wireless broadcast, so they do not shed light on our desired completion delay analysis. %
Let $2\leq J \leq P$ and $q$ be a prime power no larger than $1/p_0$. Consider a $PL\times JL$ matrix $\mathbf{F}_{J,\mathbf{\Gamma}} = [\mathbf{\Gamma}_{jj'}]_{1\leq j \leq J,1\leq j' \leq P}$ over GF($2$) with $\mathbf{\Gamma}_{jj'} \in \mathcal{C}$ subject to
\begin{itemize}
\item the $PL\times (J-1)L$ submatrix $[\mathbf{\Gamma}_{jj'}]_{1\leq j < J,1\leq j' \leq P}$ is full rank (over GF($2$)),
\item every $\mathbf{\Gamma}_{Jj'}$ is randomly selected from $\mathcal{C}$ according to the distribution prescribed in (\ref{eqn: probability distribution of coding coefficents}),
\end{itemize}
and another $P\times J$ matrix $\mathbf{F}_{J,\gamma} = [\gamma_{jj'}]_{1\leq j \leq J,1\leq j' \leq P}$ over GF($q$) with $\mathbf{\gamma}_{jj'} \in \mathrm{GF}(q)$ subject to
\begin{itemize}
\item the $P\times (J-1)$ submatrix $[\gamma_{jj'}]_{1\leq j < J,1\leq j' \leq P}$ is full rank,
\item every $\gamma_{Jj'}$ is independently and randomly selected from $\mathrm{GF(q)}$ with $\mathrm{Pr}(\gamma_{Jj'} = \gamma) = 1/q$ for all $\gamma \in \mathrm{GF}(q)$.
\end{itemize}

\begin{lemma}
\label{lemma:no_matrices_circular_shift_RLNC}
$\mathrm{Pr}(\mathrm{rank}(\mathbf{F}_{J,\mathbf{\Gamma}}) = JL) \geq 1 - p_0^{P-J+1} \geq \mathrm{Pr}(\mathrm{rank}(\mathbf{F}_{J,\gamma}) = J)$.
\begin{IEEEproof}
Please refer to Appendix-\ref{appendix:lemma_proof_matrix_number}.
\end{IEEEproof}
\end{lemma}

 Consider the proposed circular-shift RLNC scheme with $p_0 \geq 1/(L+2)$ and the conventional RLNC scheme over GF($q$) with $q$ a prime power no larger than $1/p_0$. Denote by $D^\mathrm{circ}$ and $D^{\mathrm{GF}(q)}$ their respective completion delay, as well as by $D_r^\mathrm{circ}$ and $D_r^{\mathrm{GF}(q)}$ their respective completion delay for receiver $r$. Thus, we have
\begin{equation}
\begin{split}
D^{\mathrm{circ}} &= \max\{D_1^{\mathrm{circ}}, D_2^{\mathrm{circ}}, \ldots, D_R^{\mathrm{circ}}\} \\
D^{\mathrm{GF}(q)} &= \max\{D_1^{\mathrm{GF}(q)}, D_2^{\mathrm{GF}(q)}, \ldots, D_R^{\mathrm{GF}(q)}\}.
\end{split}
\end{equation}

\begin{theorem}
\label{thm:completion_delay_circular-shift_RLNC}
$\mathbb{E}(D^{\mathrm{circ}}) \leq \mathbb{E}(D^{\mathrm{GF}(q)})$ and $\mathbb{E}(D_r^{\mathrm{circ}}) \leq \mathbb{E}(D_r^{\mathrm{GF}(q)})$.
\begin{IEEEproof}
Similar to (\ref{eqn:expection_D_general_formula}),
\begin{equation}
\begin{split}
  \mathbb{E}[D^{\mathrm{circ}}] &= \sum\nolimits_{d\geq 0}\left(1-\prod\nolimits_{1\leq r \leq R}\mathrm{Pr}(D_r^{\mathrm{circ}} \leq d)\right), \\
  \mathbb{E}[D^{\mathrm{GF}(q)}] &= \sum\nolimits_{d\geq 0}\left(1-\prod\nolimits_{1\leq r \leq R}\mathrm{Pr}(D_r^{\mathrm{GF}(q)} \leq d)\right).
\end{split}
\end{equation}
Thus, it suffices to prove that for each receiver $r$ and $d \geq 0$, \begin{equation}
\label{eqn:Dr_circ_Dr_GFq_prob_compare}
\mathrm{Pr}(D_r^{\mathrm{circ}} \leq d) \geq \mathrm{Pr}(D_r^{\mathrm{GF}(q)} \leq d)\quad \forall d \geq 0,
\end{equation}
whose proof can be found in Appendix-\ref{appendix:thm_proof_completion_delay_circular-shift_RLNC}. Since $\mathbb{E}(D_r^{\mathrm{circ}}) = \sum_{d\geq 0}\left(1 - \mathrm{Pr}(D_r^{\mathrm{circ}} \leq d)\right)$, $\mathbb{E}(D_r^{\mathrm{GF}(q)}) = \sum_{d\geq 0}\left(1 - \mathrm{Pr}(D_r^{\mathrm{GF}(q)} \leq d)\right)$, Eq. (\ref{eqn:Dr_circ_Dr_GFq_prob_compare}) implies $\mathbb{E}(D_r^{\mathrm{circ}}) \leq \mathbb{E}(D_r^{\mathrm{GF}(q)})$. %
\end{IEEEproof}
\end{theorem}

The next corollary is a direct consequence of Theorem \ref{thm:GF2_scheme_convergence} and Theorem \ref{thm:completion_delay_circular-shift_RLNC}.

\begin{corollary}
\label{corollary:circ_RLNC_convergence}
When $p_0 \leq 1/2$, $\mathbb{E}(D^{\mathrm{circ}}) \leq \mathbb{E}(D^{\mathrm{GF}(2)})$ and $\lim_{P\rightarrow\infty}\mathbb{E}[D^{\mathrm{circ}}]/P = \lim_{P\rightarrow\infty}\mathbb{E}[D^{\mathrm{perf}}]/P$.
\end{corollary}

In the proposed RLNC scheme, the set $\mathcal{C}$ of $L \times L$ matrices for coding coefficient selection is particularly designed, so that the decoding complexity of the proposed scheme can be significantly reduced compared with the conventional RLNC scheme over GF($2^L$). The insights are as follows.

First, for every symbol $\mathbf{s}$, \emph{i.e.,} a binary row vector of length $L$, when it is (right) multiplied by a binary matrix of size $L$, $L(L-1)$ binary operations are required in general. However, for a nonzero matrix $\mathbf{\Gamma} = \mathbf{G}\mathbf{C}_{L+1}^l\mathbf{H}$ in $\mathcal{C}$, it only requires $L-1$ binary operations to compute $\mathbf{s}\mathbf{\Gamma}$, where the only binary operations are performed in computing $\mathbf{s}\mathbf{G}$ while the complexity of a circular-shift operation on a binary sequence can be ignored (See, e.g., \cite{Hou2016,Sun_TIT19}). Moreover, the next proposition justifies that to compute $\mathbf{s}\mathbf{\Gamma}^{-1}$ only takes $L-1$ binary operations as well.

\begin{proposition}
\label{Prop:Gamma_property}
For a nonzero matrix $\mathbf{\Gamma} = \mathbf{G}\mathbf{C}_{L+1}^l\mathbf{H}$ in $\mathcal{C}$,
\begin{align}
\label{eqn:Gamma_times_H_expression}
\mathbf{\Gamma}\mathbf{G} &= \mathbf{G}\mathbf{C}_{L+1}^{l}, \\
\label{eqn:Gamma_inverse}
\mathbf{\Gamma}^{-1} &= \mathbf{G}\mathbf{C}_{L+1}^{L-l+1}\mathbf{H}.
\end{align}
\begin{IEEEproof}
Observe that $\mathbf{H}\mathbf{G} = \begin{bmatrix} \mathbf{I}_{L} & \mathbf{1} \\ \mathbf{0} & 0 \end{bmatrix} = \mathbf{I}_{L+1} + [\mathbf{0} ~ \mathbf{1}]$, where $[\mathbf{0} ~ \mathbf{1}]$ refers to $(L+1)\times(L+1)$ matrix with all-zero entries in the first $L$ columns and all-one entries in the last column. %
Consequently, $(\mathbf{G}\mathbf{C}_{L+1}^l\mathbf{H})\mathbf{G} = \mathbf{I}_L + \mathbf{G}\mathbf{C}_{L+1}^l[\mathbf{0} ~ \mathbf{1}]$. Because $\mathbf{C}_{L+1}^l[\mathbf{0} ~ \mathbf{1}] = [\mathbf{0} ~ \mathbf{1}]$ and $\mathbf{G}[\mathbf{0} ~ \mathbf{1}] = \mathbf{0}$, $\mathbf{G}\mathbf{C}_{L+1}^l[\mathbf{0} ~ \mathbf{1}] = \mathbf{0}$.  Eq. (\ref{eqn:Gamma_times_H_expression}) is thus proved. %
Due to (\ref{eqn:Gamma_times_H_expression}), $(\mathbf{G}\mathbf{C}_{L+1}^l\mathbf{H})(\mathbf{G}\mathbf{C}_{L+1}^{L-l+1}\mathbf{H}) = \mathbf{G}\mathbf{C}_{L+1}^l\mathbf{C}_{L+1}^{L-l+1}\mathbf{H} = \mathbf{G}\mathbf{I}_{L+1}\mathbf{H} = \mathbf{I}_L$.
\end{IEEEproof}
\end{proposition}

Moreover, for a positive integer $J$, consider a full rank (over GF($2$)) block matrix $[\mathbf{\Gamma}_{jj'}]_{1 \leq j,j' \leq J}$ with every nonzero block $\mathbf{\Gamma}_{jj'} = \mathbf{G}\mathbf{C}_{L+1}^{a_{jj'}}\mathbf{H}$. We next provide a formula to concisely characterize the inverse matrix $\mathbf{D} = [\mathbf{B}_{jj'}]_{1\leq j,j' \leq J}$ of $[\mathbf{\Gamma}_{jj'}]_{1 \leq j,j' \leq J}$, where every block $\mathbf{B}_{jj'}$ is also an $L\times L$ matrix. It turns out that for every block $\mathbf{B}_{jj'}$ in $\mathbf{D}$, to compute $\mathbf{s}\mathbf{B}_{jj'}$ requires at most $L(L+1)/2 - 2$ binary operations, which are also fewer than the number $L(L-1)$ of binary operations to compute $\mathbf{s}$ multiplied by a general matrix.

Observe that the set $\mathcal{R} = \{\sum_{l=0}^{L}a_l\mathbf{C}_{L+1}^l, a_l \in \mathrm{GF}(2)\}$ forms a commutative ring of circulant matrices of size $L+1$. In $\mathcal{R}$, for every $\mathbf{\Psi} = \sum_{l=0}^{L}a_l\mathbf{C}_{L+1}^l$, define $\sigma(\mathbf{\Psi})$ of it as follows. If the number of nonzero coefficients $\{a_0, a_1, \ldots, a_L\}$ is larger than $L/2$, then $\sigma(\mathbf{\Psi}) = \sum_{l=0}^{L}(1+a_l)\mathbf{C}_{L+1}^l$, \emph{i.e.}, $\sigma(\mathbf{\Psi}) = \mathbf{\Psi}+\mathbf{1}$. Otherwise, $\sigma(\mathbf{\Psi}) = \mathbf{\Psi}$. In this way, $\sigma(\mathbf{\Psi})$ is always a summation of at most $L/2$ cyclic permutation matrices. %

In addition, for every matrix $[\mathbf{\Psi}_{jj'}]_{1 \leq j,j' \leq J}$ over $\mathcal{R}$, though it can be regarded as a matrix of size $J(L+1)$ over GF($2$), we impose the computation of its determinant to be conducted over $\mathcal{R}$. Specifically, denote by $\mathbf{\Lambda}$ the determinant of $\left([\mathbf{\Psi}_{jj'}]_{1 \leq j,j' \leq J}\right)$ over $\mathcal{R}$
\begin{equation}
\label{eqn:determinant_block}
\mathbf{\Lambda} = \det\left([\mathbf{\Psi}_{jj'}]_{1 \leq j,j' \leq J}\right)
= \sum\nolimits_{\tau \in S_J} \prod\nolimits_{j = 1}^{J} \mathbf{\Psi}_{j\tau(j)},
\end{equation}
where $S_J$ refers to the permutation group consisting of all permutations on $\{1, 2, \ldots, J\}$.

\begin{theorem}
\label{thm:Gamma_matrix_inverse}
Consider a full rank matrix $[\mathbf{G}\mathbf{\Psi}_{jj'}\mathbf{H}]_{1 \leq j,j' \leq J}$, where $\mathbf{\Psi}_{jj'} \in \mathcal{R}$, its inverse matrix can be represented as the block matrix $\mathbf{D} = [\mathbf{B}_{jj'}]_{1\leq j,j' \leq J}$ with every block of $L \times L$ matrix $\mathbf{B}_{jj'}$ defined as
\begin{equation}
\label{eqn:decoding_matrix_block_formula}
\mathbf{G} \sigma(\mathbf{\Lambda}^{2^L-2}\det(\mathbf{M}_{jj'})) \mathbf{H},
\end{equation}
where $\mathbf{M}_{jj'}$ is the matrix obtained from $[\mathbf{\Psi}_{jj'}]_{1 \leq j,j' \leq J}$ by deleting its ${j'}^\mathrm{th}$ block row and ${j}^\mathrm{th}$ block column, and its determinant is also computed over $\mathcal{R}$.
\begin{IEEEproof}
For $1 \leq j, j' \leq J$, denote by $\mathbf{B}'_{jj'} \in \mathcal{R}$ the $(L+1)\times (L+1)$ matrix
\begin{equation}
\mathbf{B}'_{jj'} = \mathbf{\Lambda}^{2^L-2}\det(\mathbf{M}_{jj'}),
\end{equation}
so that it is equivalent to show
\begin{equation}
\label{eqn:equivalent_decoding_matrix_block_formula}
[\mathbf{G}\mathbf{\Psi}_{jj'}\mathbf{H}]_{1 \leq j,j' \leq J}[\mathbf{G}\sigma(\mathbf{B}'_{jj'})\mathbf{H}]_{1 \leq j,j' \leq J} = \mathbf{I}_{JL}.
\end{equation} %
By application of Lemma 3 in \cite{Sun_letter19}, we can obtain the following proposition
\begin{equation}
\label{eqn:inverse_from_comm_letter_paper}
\begin{split}
[ \mathbf{G}\mathbf{\Psi}_{jj'}]_{1 \leq j,j' \leq J}[\mathbf{B}'_{jj'} \mathbf{H}]_{1 \leq j,j' \leq J} = \mathbf{I}_{JL}.
\end{split}
\end{equation}
Further, since Eq. (\ref{eqn:Gamma_times_H_expression}) in Proposition \ref{Prop:Gamma_property} implies that for $\mathbf{\Psi}_1, \mathbf{\Psi}_2 \in \mathcal{R}$,
\begin{equation}
(\mathbf{G}\mathbf{\Psi}_1\mathbf{H})(\mathbf{G}\mathbf{\Psi}_2\mathbf{H}) = \mathbf{G}\mathbf{\Psi}_1\mathbf{\Psi}_2\mathbf{H},
\end{equation}
the following can be deduced from Eq. (\ref{eqn:inverse_from_comm_letter_paper})
\begin{equation}
\begin{split}
[\mathbf{G}\mathbf{\Psi}_{jj'}\mathbf{H}]_{1 \leq j,j' \leq J}[\mathbf{G}\mathbf{B}'_{jj'}\mathbf{H}]_{1 \leq j,j' \leq J}  = \mathbf{I}_{JL}.
\end{split}
\end{equation}
By the definition of $\sigma$, either $\sigma(\mathbf{B}'_{jj'}) = \mathbf{B}'_{jj'}$ or $\sigma(\mathbf{B}'_{jj'}) = \mathbf{B}'_{jj'} + \mathbf{1}$. Since $\mathbf{G}\mathbf{1} = \mathbf{0}$,
\begin{equation}
[\mathbf{G}\mathbf{\Psi}_{jj'}\mathbf{H}]_{1 \leq j,j' \leq J} [\mathbf{G}\sigma(\mathbf{B}'_{jj'})\mathbf{H}]_{1 \leq j,j' \leq J} = [\mathbf{G}\mathbf{\Psi}_{jj'}\mathbf{H}]_{1 \leq j,j' \leq J} [\mathbf{G}\mathbf{B}'_{jj'}\mathbf{H}]_{1 \leq j,j' \leq J}  = \mathbf{I}_{JL}.
\end{equation}
\end{IEEEproof}
\end{theorem}

\begin{example}
Assume $[\mathbf{\Psi}_{jj'}]_{1 \leq j,j' \leq 3} = \begin{bmatrix} \mathbf{I}_5 & \mathbf{C}_5 & \mathbf{C}_5 \\
\mathbf{I}_5 & \mathbf{C}_5^2 & \mathbf{C}_5^3 \\
\mathbf{I}_5 & \mathbf{C}_5^3 & \mathbf{C}_5^4 \end{bmatrix}$. One may check that $\det([\mathbf{\Psi}_{jj'}]_{1 \leq j,j' \leq 3}) = \mathbf{I}_5+\mathbf{C}_5^3$, and %
%$[\mathbf{M}_{jj'}]_{1 \leq j, j' \leq 3} = \begin{bmatrix} \mathbf{0} & \mathbf{I} + \mathbf{C}_5^4 & \mathbf{C}_5^3+\mathbf{C}_5^4 \\
%\mathbf{C}_5^3+\mathbf{C}_5^4 & \mathbf{C}_5+\mathbf{C}_5^4 & \mathbf{C}_5+\mathbf{C}_5^3 \\
%\mathbf{C}_5^2+\mathbf{C}_5^3 & \mathbf{C}_5+\mathbf{C}_5^3 & \mathbf{C}+\mathbf{C}_5^2 \end{bmatrix}$. %
$[\mathbf{M}_{jj'}]_{1 \leq j, j' \leq 3} = \begin{bmatrix} \mathbf{0} & \mathbf{I} + \mathbf{C}_5^4 & \mathbf{C}_5^3+\mathbf{C}_5^4 \\
\mathbf{C}_5^3+\mathbf{C}_5^4 & \mathbf{C}_5+\mathbf{C}_5^4 & \mathbf{C}_5+\mathbf{C}_5^3 \\
\mathbf{C}_5^2+\mathbf{C}_5^3 & \mathbf{C}_5+\mathbf{C}_5^3 & \mathbf{C}+\mathbf{C}_5^2 \end{bmatrix}$. %
Based on  (\ref{eqn:decoding_matrix_block_formula}), %
$[\mathbf{B}_{jj'}]_{1\leq j, j' \leq 3} = (\mathbf{I}_3\otimes \mathbf{G})\begin{bmatrix} \mathbf{0} & \mathbf{C}_5^2 + \mathbf{C}_5^4 & \mathbf{C}_5+\mathbf{C}_5^3 \\
\mathbf{C}_5+\mathbf{C}_5^3 & \mathbf{C}_5 & \mathbf{C}_5^3 \\
\mathbf{I}_5+\mathbf{C}_5^2 & \mathbf{C}_5^3 & \mathbf{C}_5+\mathbf{C}_5^4 \end{bmatrix} (\mathbf{I}_3\otimes \mathbf{H})$, where $\otimes$ means the Kronecker product. %
One may further check that $(\mathbf{I}_3\otimes\mathbf{G})[\mathbf{\Psi}_{jj'}]_{1 \leq j,j' \leq 3} (\mathbf{I}_3\otimes\mathbf{H})[\mathbf{B}_{jj'}]_{1\leq j, j' \leq 3} = \mathbf{I}_{12}$. \hfill $\blacksquare$ %
\end{example}

\begin{remark}
Compared with the circular-shift LNC schemes considered in \cite{Sun_TIT19}, the one proposed in this paper does not introduce any redundant bit for transmission. The key is Eq. (\ref{eqn:Gamma_times_H_expression}) and (\ref{eqn:Gamma_inverse}) in Proposition \ref{Prop:Gamma_property}, which guarantees the correctness of Theorem \ref{thm:Gamma_matrix_inverse}. The insight brought about from  (\ref{eqn:Gamma_times_H_expression}) and (\ref{eqn:Gamma_inverse}) is that for the (wireline) networks as modeled in \cite{Sun_TIT19}, every circular-shift LNC scheme with redundant bits for transmission may be theoretically transformed to a vector LNC scheme without redundant bits and with the coding coefficients selected from $\mathcal{R}$. This topic is beyond the scope of this paper and will be investigated elsewhere.
\end{remark}

A detailed discussion on the decoding process and the decoding complexity are given in next two subsections.

\subsection{Decoding Algorithm}

Assume that receiver $r$ has successfully obtained $P$ packets $\mathbf{m}_j' = \sum_{1\leq j'\leq P} \mathbf{\Gamma}_{jj'}\circ \mathbf{m}_{j'}$, $1 \leq j \leq P$, such that the $PL\times PL$ matrix $[\mathbf{\Gamma}_{jj'}]_{1 \leq j,j' \leq P}$ over GF($2$) is full rank. Write every coding coefficient $\mathbf{\Gamma}_{jj'}$ in the form of $\mathbf{G}\mathbf{C}_{jj'}\mathbf{H}$, with $\mathbf{C}_{jj'}$ to be either a cyclic permutation matrix or the zero matrix of size $L+1$. Without loss of generality, assume $\mathbf{m}_1', \ldots, \mathbf{m}_{U_r}'$ are uncoded packets, and $\mathbf{m}_j' = \mathbf{m}_j$ for $1 \leq j \leq U_r$. Receiver $r$ remains to decode $P - U_r$ original packets $\mathbf{m}_{U_r+1}, \ldots, \mathbf{m}_{P}$. %

Denote by $[\mathbf{B}_{jj'}]_{1 \leq j, j' \leq P}$ the inverse matrix of $[\mathbf{\Gamma}_{jj'}]_{1 \leq j,j' \leq P}$ formulated in Theorem \ref{thm:Gamma_matrix_inverse}. Thus, for $U_r < j \leq P$, $\mathbf{m}_j = \sum\nolimits_{1 \leq j'\leq P}\mathbf{B}_{jj'}\circ\mathbf{m}_{j'}'$. %
By making use of the property that $\mathbf{\Gamma}_{jj} = \mathbf{I}_L$ and $\mathbf{\Gamma}_{jj'} = \mathbf{0}$ for $1 \leq j \leq U_r$, $j' \neq j$, we can further refine the above decoding process without directly calculating $\mathbf{B}_{jj'}$ from $[\mathbf{\Gamma}_{jj'}]_{1 \leq j,j' \leq P}$, as explicated below.

% In this subsection, for $U_r+1 \leq j \leq P$, denote by $\mathbf{F}_j$ the coding vector, which is an $PL\times L$ matrix, for the coded packet $\mathbf{m}_{j}'$.

The refined decoding process of the proposed circular-shift RLNC scheme consists of two phases. In phase I decoding, first update every received packet $\mathbf{m}_j'$, $1 \leq j \leq P$ according to
\begin{equation}
\label{eqn: G_phase_I}
\mathbf{m}_j' = \mathbf{G}\circ\mathbf{m}_j',
\end{equation}
where $\circ$ is defined in (\ref{eqn:circ-symbol-definition}), so that each of the $\frac{M}{L}$ symbols in $\mathbf{m}_j'$ has length $L+1$. %
For $U_r < j \leq P$, $\mathbf{m}_{j}'$ is further updated as
\begin{equation}
\label{eqn: procedure phase I_circular_shift}
\mathbf{m}_j' = \mathbf{m}_j' - \sum\nolimits_{1 \leq j' \leq U_r} \mathbf{C}_{jj'} \circ \mathbf{m}_{j'}'.
\end{equation}
Slightly different from (\ref{eqn:circ-symbol-definition}), the operator $\circ$ in (\ref{eqn: procedure phase I_circular_shift}) works for row vectors of $\frac{M}{L}$ symbols with symbol length $L+1$. %
Note that every updated coded packet $\mathbf{m}_{j}'$, $U_r < j \leq P$, can be represented as $\mathbf{m}_{j}' = \sum_{U_r < j' \leq P} (\mathbf{\Gamma}_{jj'}\mathbf{G})\circ\mathbf{m}_{j'} = \sum_{U_r < j' \leq P}(\mathbf{G}\mathbf{C}_{jj'})\circ\mathbf{m}_{j'}$, where the last equality holds due to (\ref{eqn:Gamma_times_H_expression}). In addition, the $(P-U_r)L\times(P-U_r)L$ matrix $[\mathbf{\Gamma}_{jj'}]_{U_r < j,j' \leq P} (= [\mathbf{G}\mathbf{C}_{jj'}\mathbf{H}]_{U_r < j,j' \leq P})$ is full rank. Write $\mathcal{A} = \mathcal{A}' = \{U_r+1, U_r+2, \ldots, P\}$.

In the first step of phase II decoding, recursively perform the following procedure. From $\mathcal{A}$, select an index, say $j_0$, such that there is only one nonzero coding coefficient among $\{\mathbf{\Gamma}_{j_0j'}: j' \in \mathcal{A}'\}$. Let $\mathbf{\Gamma}_{j_0j_0'} (= \mathbf{G}\mathbf{C}_{j_0j_0'}\mathbf{H})$ denote this nonzero coefficient block. Thus, $\mathbf{m}_{j_0}' = (\mathbf{\Gamma}_{j_0j_0'}\mathbf{G}) \circ \mathbf{m}_{j_0'} =  (\mathbf{G}\mathbf{C}_{j_0j_0'}) \circ \mathbf{m}_{j_0'}$. Remove $j_0$ and $j_0'$ from $\mathcal{A}$ and $\mathcal{A}'$ respectively, and reset
\begin{equation}
\label{eqn:recover_packet_by_Gamma_inverse}
\mathbf{m}_{j_0}' = \mathbf{C}_{j_0j_0'}^{-1} \circ \mathbf{m}_{j_0}'.
\end{equation}
The original packet $\mathbf{m}_{j_0'}$ is then equal to $\mathbf{H}\circ\mathbf{m}_{j_0}'$. Corresponding to every remaining index $j$ in $\mathcal{A}$, reset the coded packet $\mathbf{m}_{j}'$ by
\begin{equation}
\label{eqn: procedure phase II A_circular_shift}
\mathbf{m}_{j}' = \mathbf{m}_{j}' - \mathbf{C}_{jj_0'}\circ\mathbf{m}_{j_0'},
\end{equation}

The above recursive procedure ends till for every $j \in \mathcal{A}$, there are at least two nonzero coding coefficient blocks among $\{\mathbf{\Gamma}_{jj'}: j' \in \mathcal{A}'\}$.

After the first step of phase II decoding, if $\mathcal{A}$ is empty, then all $P$ original packets have been successfully recovered. Otherwise, $|\mathcal{A}| = |\mathcal{A}'| \geq 2$, and $\mathbf{m}_j' =  \sum_{j'\in \mathcal{A}'} (\mathbf{\Gamma}_{jj'}\mathbf{G})\circ\mathbf{m}_{j'}$ for $j \in \mathcal{A}$. Continue to perform the second step as follows. %
First, arbitrarily find an index $j_1$ in $\mathcal{A}$, and a nonzero coefficient, say $\mathbf{\Gamma}_{j_1j_1'} (= \mathbf{G}\mathbf{C}_{j_1j_1'}\mathbf{H})$, in $\{\mathbf{\Gamma}_{j_1j'}: j' \in \mathcal{A}'\}$. Reset
\begin{equation}
\label{eqn:inverse of a nonzero coefficient}
\mathbf{m}_{j_1}' = \mathbf{C}_{j_1j_1'}^{-1}\circ\mathbf{m}_{j_1}'.
\end{equation}
Next, remove $j_1$ and $j_1'$ from $\mathcal{A}$ and $\mathcal{A}'$ respectively. For every $j \in \mathcal{A}$, $j' \in \mathcal{A}'$, update $\mathbf{m}_{j}'$ and $\mathbf{\Gamma}_{jj'}$ as
\begin{equation}
\label{eqn:turn all elements to zero except one}
\mathbf{m}_{j}' = \mathbf{m}_{j}' - \mathbf{C}_{jj_1'}\circ\mathbf{m}_{j_1}',
\mathbf{\Gamma}_{jj'} = \mathbf{G}(\mathbf{C}_{jj'}+\mathbf{C}_{j_1j'}\mathbf{C}_{j_1j_1'}^{-1})\mathbf{H}.
\end{equation}
After the update, $\mathbf{m}_{j}'$ keeps equal to $\sum_{j'\in \mathcal{A}'} (\mathbf{\Gamma}_{jj'}\mathbf{G})\circ \mathbf{m}_{j'}$. Since $\mathbf{C}_{jj'}+\mathbf{C}_{j_1j'}\mathbf{C}_{j_1j_1'}^{-1} \in \mathcal{R}$, the inverse matrix of $[\mathbf{\Gamma}_{jj'}]_{j\in \mathcal{A},j' \in \mathcal{A}'}$ can be computed based on Theorem \ref{thm:Gamma_matrix_inverse}. %
Following (\ref{eqn:decoding_matrix_block_formula}), express this inverse matrix as $[\mathbf{G}\mathbf{\Phi}_{j'j}\mathbf{H}]_{j' \in \mathcal{A}',j\in \mathcal{A}}$, where the columns and rows are respectively indexed by $\mathcal{A}'$ and $\mathcal{A}$, and every $\mathbf{\Phi}_{j'j} \in \mathcal{R}$ is a summation of at most $L/2$ cyclic permutation matrices. For $j' \in \mathcal{A}'$, compute
\begin{equation}
\label{eqn: procedure phase II B_circular_shift}
\tilde{\mathbf{m}}_{j'} = \sum\nolimits_{j \in \mathcal{A}} \mathbf{\Phi}_{j'j}\circ\mathbf{m}_{j}',
\end{equation}
and recover the original packet $\mathbf{m}_{j'}$ by setting $\mathbf{m}_{j'} = \mathbf{H}\circ\tilde{\mathbf{m}}_{j'}$.
Finally, $\mathbf{m}_{j_1'}$ can be recovered via
\begin{equation}
\label{eqn:back substitution}
\mathbf{m}_{j_1'} = \mathbf{H}\circ\left(\mathbf{m}'_{j_1} - \sum\nolimits_{j'\in \mathcal{A}'} (\mathbf{C}_{j_1j'}\mathbf{C}_{j_1j_1'}^{-1})\circ\tilde{\mathbf{m}}_{j'}\right)
\end{equation}

\subsection{Decoding Complexity Analysis}

We now analyze the number of binary operations involved in the refined decoding process described in the above subsection.
The number of binary operations required in (\ref{eqn: G_phase_I}) to update packet $\mathbf{m}_{j}'$ for all $1 \leq j \leq P$ is

\begin{equation}
\label{eqn: number of binary operations in sG}
P\frac{M}{L}(L-1)
\end{equation}
 After the update, each of the $\frac{M}{L}$ symbols in $\mathbf{m}_j'$ has length $L+1$. Since the symbol-wise multiplication $\mathbf{C}_{L+1}^l\circ \mathbf{m}_j'$ involves no binary operations, and the nonzero probability for $\mathbf{C}_{jj'}$ is $1 - p_0$, the number of binary operations required at a receiver for (\ref{eqn: procedure phase I_circular_shift}) in phase I is
\begin{equation}
\label{eqn: number of binary operations in phase I C-S}
(P-U_r)U_r(1-p_0)\frac{M}{L}(L+1).
\end{equation}

In the first step of phase II decoding, the number of required binary operations is
\begin{equation}
\label{eqn: number of binary operations in phase II A C-S}
\sum\nolimits_{j=|\mathcal{A}|+1}^{P-U_r}\frac{M}{L}(j-1)(1-p_0)(L+1).
\end{equation}
where $\mathcal{A}$ refers to the index set defined at the end of the first step. %
In the second step of phase II decoding, it takes negligible binary operations to compute (\ref{eqn:inverse of a nonzero coefficient}), and $(|\mathcal{A}|-1)(L+1)\frac{M}{L}$ binary operations to update $\mathbf{m}_{j}'$ in (\ref{eqn:turn all elements to zero except one}). %
Moreover, in (\ref{eqn: procedure phase II B_circular_shift}), every block $\mathbf{\Phi}_{j'j} \in \mathcal{R}$ can be expressed as $\sigma(\mathbf{\Lambda}^{2^L-2}\det(\mathbf{M}_{jj'}))$ according to Theorem \ref{thm:Gamma_matrix_inverse} and is a summation of at most $L/2$ cyclic permutation matrices. %
Thus, it takes at most $\frac{M}{L}(\frac{L}{2}-1)(L+1)$ binary operations to compute one term $\mathbf{\Phi}_{j'j}\circ\mathbf{m}_{j'}$ in (\ref{eqn: procedure phase II B_circular_shift}). %
It takes each receiver at most
\begin{equation}
\label{eqn: number of binary operations in phase II B C-S}
(|\mathcal{A}|-1)^2(\frac{L}{2}-1)\frac{M}{L}(L+1)+(|\mathcal{A}|-1)(|\mathcal{A}|-2)\frac{M}{L}(L+1)
\end{equation}
binary operations to compute (\ref{eqn: procedure phase II B_circular_shift}). %
Note that same as in the analysis of conventional RLNC schemes, we ignore the number of operations involved in obtaining $[\mathbf{G}\mathbf{\Phi}_{j'j}\mathbf{H}]_{j' \in \mathcal{A}',j\in \mathcal{A}}$, since $|\mathcal{A}|$ is relatively small. %
The number of binary operations in the final procedure (\ref{eqn:back substitution}) to recover the original packet $\mathbf{m}_{P-|\mathcal{A}|+j_0}$ is $M/L(|\mathcal{A}|-1)(1-p_0)(L+1)$, where $1 - p_0$ represents the probability for every term $\mathbf{C}_{j_{1}j'}$ in (\ref{eqn:back substitution}) to be nonzero.

As $\mathbb{E}[U_r]=Pp_r$ and $\mathbb{E}[U_r^2] = Pp_r(Pp_r - p_r +1)$, the expected number of binary operations for the whole decoding process is

\begin{equation}
\label{eqn:total_decoding_complexity_C-S}
\begin{split}
\frac{M}{L}[P(L-1)+(L+1)(P^2p_r(1-p_0)(1-p_r)+(|\mathcal{A}|-1)^2(\frac{L}{2}-1)\\
+(|\mathcal{A}|-1)(|\mathcal{A}|-p_0))]+\sum\nolimits_{j=|\mathcal{A}|+1}^{P-U_r}\frac{M}{L}(j-1)(1-p_0)(L+1).
\end{split}
\end{equation}

\begin{remark}
If we allow one bit redundancy per symbol for transmission, then the proposed circular-shift RLNC scheme can be modified by randomly selecting the nonzero coding coefficients from the set $\{\mathbf{G}\mathbf{C}_{L+1}, \mathbf{G}\mathbf{C}_{L+1}^2, \ldots, \mathbf{G}\mathbf{C}_{L+1}^{L+1}\}$ instead of from $\mathcal{C}$. Once the coding coefficients have been randomly selected, the formed code essentially belongs to the same type of circular-shift linear network codes with $1$-bit redundancy originally proposed in \cite{Sun_TIT19} for wireline networks. In this way, every packet transmitted from the source in the modified scheme still consists $\frac{M}{L}$ symbols, but every symbol contains $L+1$ bits. Based on (\ref{eqn:Gamma_times_H_expression}), every packet transmitted by the modified scheme can be expressed as $\sum\nolimits_{j=1}^{P}(\mathbf{\Gamma}_j\mathbf{G}) \circ \mathbf{m}_j$. As a consequence, what the receiver received in the modified scheme are directly the packets obtained after (\ref{eqn: G_phase_I}). The decoding process described in the previous subsection can thus be slightly simplified so that a total of $P\frac{M}{L}(L-1)$ binary operations (to perform (\ref{eqn: G_phase_I})) can be saved at decoding. This is at a cost of $\frac{M}{L}$ extra bits per packet to be transmitted.
\end{remark}

\section{Numerical Analysis}
In this section, we present numerical results to compare the performance of the proposed RLNC schemes with conventional RLNC schemes. Note that even though IDNC is also a type of network coding schemes designed for wireless broadcast, it requires feedback from every receiver after every packet broadcast to indicate whether the packet is received, and hence has a different network assumption from ours. As a result, in the present paper, same as in other studies \cite{Yu&Parastoo2018,Ioannis2017,Atilla2013,Emmanouil2016,NanXie2013,SkevakisL2016,SkevakisL2017,Lucani2012,Tsimbalo2018} about RLNC in wireless broadcast, we shall not compare the performance of our proposed schemes with IDNC. We assume that the system has $60$ receivers and in each round of successful transmission of $P$ packets to all receivers, the packet erasure probability per receiver is a fixed value uniformly selected between $0.1$ and $0.2$. In the figure legend, the conventional RLNC scheme over GF($2^L$) is labeled as ``GF($2^\mathrm{L}$)'', the proposed circular-shift RLNC scheme with zero entry probability $p_0$ is labeled as ``C-S $p_0$-0'', and the circular-shift RLNC scheme with $1$-bit redundancy per symbol is labeled as ``C-S.R $p_0$-0''.

\begin{figure}[t!]
\centering
\includegraphics[width=0.7\linewidth]{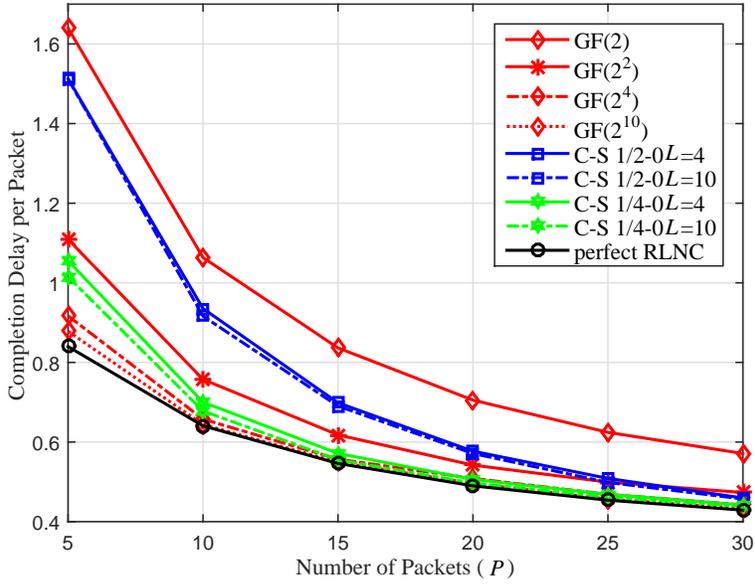}
%\scalebox{0.66}
%{\includegraphics{Fig1.eps}}
\caption{Average completion delay per packet.}
\label{fig:Completion delay per packet versus number of packets P}
\end{figure}

\begin{figure}[!t]
\setlength{\abovecaptionskip}{0.1cm}
\setlength{\belowcaptionskip}{-0.cm}
\centering
\subfigure[]{
\includegraphics[width=0.7\linewidth]{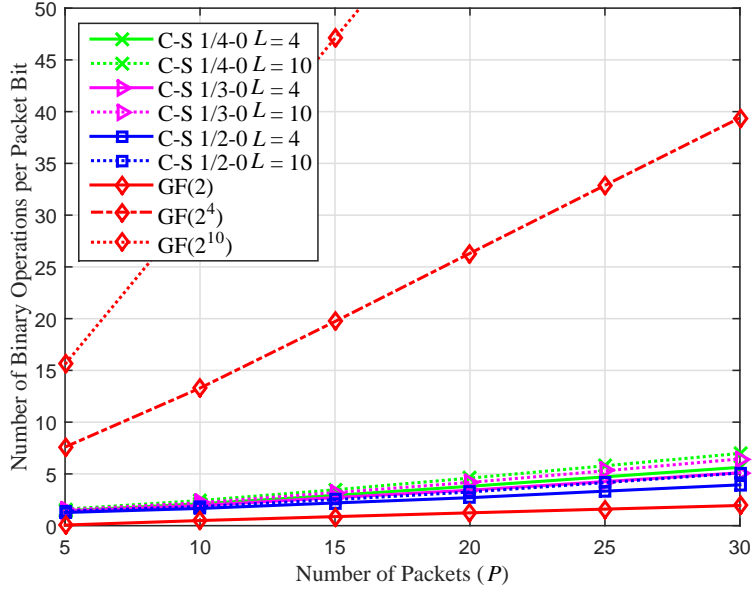}
%\caption{fig1}
}
\subfigure[]{
\includegraphics[width=0.7\linewidth]{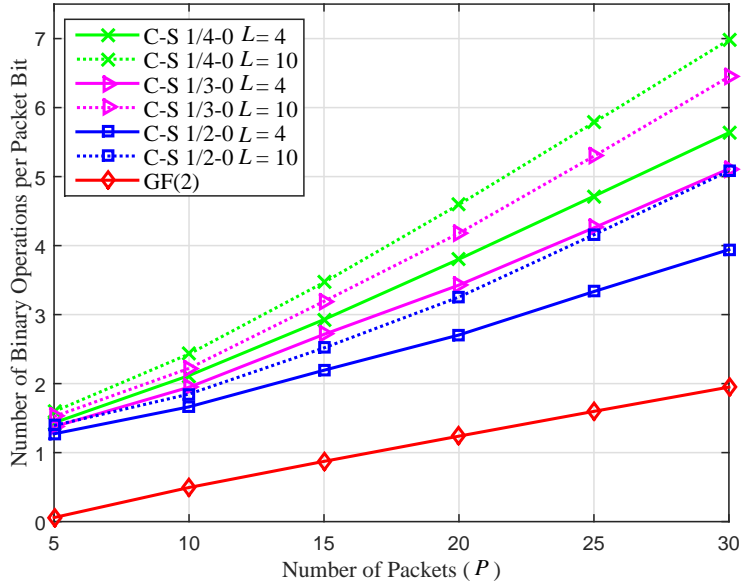}
}
\caption{Average number of binary operations for decoding per packet bit.}
\label{fig:Decoding_complexity}
\end{figure}

Fig. \ref{fig:Completion delay per packet versus number of packets P} compares the average completion delay per packet of different RLNC schemes as a function of the number of packets $P$. One may observe the followings from this figure. First, the average completion delay of every RLNC scheme, including the conventional one over GF($2$), converges to the one of perfect RLNC, which means linear independence among arbitrary $P$ packets generated by the source, and thus has the smallest completion delay. This result is in line with Theorem \ref{thm:GF2_scheme_convergence} and Corollary \ref{corollary:circ_RLNC_convergence}. %
In particular, regardless of the choice of $L$ and $p_0$, the convergence rates of our proposed scheme is higher than that of the conventional RLNC scheme over GF($2$), and the average completion delay of the proposed scheme is always smaller than that of the conventional RLNC scheme over GF($2$). This is justified by Theorem \ref{thm:completion_delay_circular-shift_RLNC}. %
 Second, when $p_0 = 1/4$, the average completion delays of the proposed schemes for both $L = 4$ and $L = 10$ are smaller than that of the conventional RLNC scheme over GF($2^2$) too. This observation numerically verifies Theorem \ref{thm:completion_delay_circular-shift_RLNC} once more and demonstrates that the theoretical upper bound in Theorem \ref{thm:completion_delay_circular-shift_RLNC} becomes tighter with the decrease of $p_0$. %
Third, for fixed $L$, the completion delays of the proposed schemes are lower bounded by the conventional RLNC schemes over GF($2^L$), and the gap is vanishing with the increase of $P$ and the decrease of $p_0$. In particular, when $P \geq 15$ and $p_0= 1/4$, the gap becomes negligible. %
Last, with the decrease of $p_0$ or the increase of $L$, the average completion delay can be reduced, but the benefit from the latter is not as obvious as the former.

Fig. \ref{fig:Decoding_complexity}.(a) and (b) shows the decoding complexity performance of different RLNC schemes as a function of $P$. The decoding complexity is measured by the average number of binary operations required in the decoding process described in Sec. II-C and Sec. III-B. %
Specifically, the decoding complexities for RLNC schemes over GF($2^4$) and GF($2^{10}$) are based on (\ref{general eqn: complexity of phase I}), (\ref{general complete eqn: complexity of phase II A}) and (\ref{general eqn: complexity of phase II B}), the one for the RLNC scheme over GF($2$) is based on (\ref{eqn:decoding_complexity_GF2}), and the ones for the proposed RLNC schemes are based on (\ref{eqn:total_decoding_complexity_C-S}), all with the average value for the variable $U_r$ and $|\mathcal{A}|$ obtained via simulation. The average number of binary operations for decoding depicted in Fig. \ref{fig:Decoding_complexity} is normalized by the packet number $P$ and the packet length $M$. %
From Fig. \ref{fig:Decoding_complexity}.(a), we can see that with increasing $P$, the decoding complexity of every RLNC scheme increases, while the increasing rates of our proposed schemes are much smaller than that of the conventional RLNC scheme over GF($2^4$) and are close to that of the conventional RLNC over GF($2$). %

Fig. \ref{fig:Decoding_complexity}.(b) depicts a clearer comparison of the decoding complexity between our proposed schemes and the RLNC scheme over GF($2$). %
It illustrates that under the same symbol length $L$, the larger the $p_0$ is, the closer the decoding complexity is to GF($2$). Note that the decoding complexity in Fig. \ref{fig:Decoding_complexity} for the RLNC scheme over GF($2$) is based on (\ref{eqn:decoding_complexity_GF2}), so it is a lower bound on the actual number of binary operations required for decoding. %
Despite of this, one may observe that for the case $L = 4$ and $p_0 = 1/4$, when $P \geq 15$, the proposed new scheme has the average binary operation number for decoding only about $3$ times the one of the RLNC scheme over GF($2$), while its the average completion delay is merely within $5\%$ higher than the perfect RLNC scheme.

\begin{figure}[t!]
\centering
\includegraphics[width=0.7\linewidth]{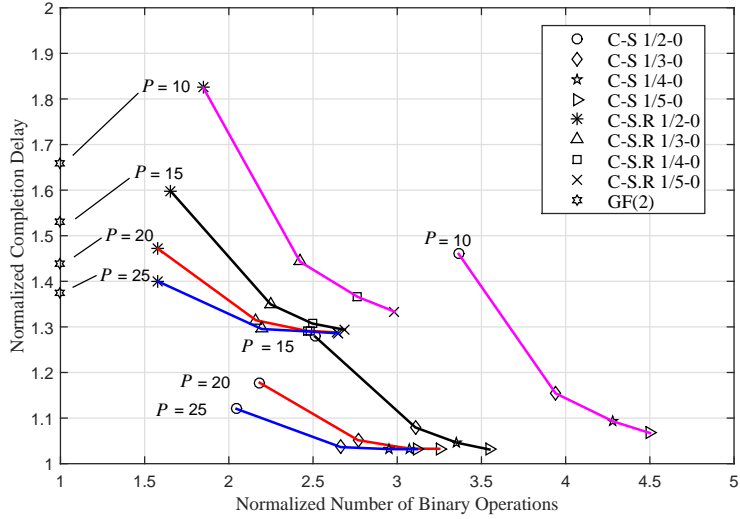}
%\scalebox{0.66}
\caption{Trade-off between the average completion delay, which is normalized by the one of perfect RLNC, and the average number of binary operations for decoding, which is normalized by the one of the RLNC scheme over GF($2$).}
\label{fig:Trade-off}
\end{figure}

Recall that in the proposed circular-shift RLNC scheme without redundancy and the counterpart scheme with $1$ bit redundancy per symbol as described in Remark 3, a smaller $p_0$ yields a higher decoding complexity but a lower average completion delay. In Fig. \ref{fig:Trade-off}, we evaluate the interesting delay-complexity trade-off for these schemes with fixed $L = 4$. The decoding complexity is again measured as the average number of binary operations for decoding, and it is normalized by the one of the RLNC scheme over GF($2$). The average completion delay is normalized by the one of the perfect RLNC. The RLNC scheme over GF($2$) is also included as the benchmark scheme. %
It can be observed that for all possible $P$, the completion delay can be reduced by increasing allowable decoding complexity in both the proposed new scheme and the counterpart scheme with redundancy, but there is such a point that the completion delay cannot benefit much more from the further increase of allowable decoding complexity. %The floor on the completion delay for the scheme with redundancy is $1/L$ higher than the one for the scheme without redundancy. %
Moreover, for relatively large $P$, the scheme without redundancy generally provides a better delay-complexity trade-off than the scheme with redundancy per symbol. For instance, when $P \geq 20$, under a same allowable decoding complexity, the completion delay of the proposed scheme without redundancy is always lower than that of the scheme with redundancy. However, for smaller $P$ and relatively small allowable decoding complexity, the scheme with redundancy in turn provides a good supplement to the delay-complexity trade-off. For instance, when $P = 10$ and the allowable decoding complexity is smaller than $3$, a scheme with redundancy can be specially designed (by appropriately choosing $p_0$) to tune the (normalized) completion delay to between $1.33$ and $1.66$, where $1.66$ reaches the one of the GF($2$)-RLNC scheme. In the case that the lowest decoding complexity is desired, the GF($2$)-RLNC scheme can be adopted directly at the cost of the highest completion delay.

\section{Concluding Remarks}
For wireless broadcast without feedback, this paper made a comprehensive study on various systematic random linear network coding (RLNC) schemes from the perspective of completion delay and decoding complexity. In particular, the proposed method to design RLNC schemes is based on circular-shift operations and a new parameter to control the zero-coding coefficient probability, and it presents itself as a new paradigm for RLNC design to have a much better trade-off between completion delay and decoding complexity. %
As potential future works, it is interesting to rigorously define a metric on the trade-off between the completion delay and decoding complexity, obtain a tighter upper bound on the average completion delay of the proposed RLNC schemes, and explore the applicability of the proposed RLNC schemes to deadline constrained wireless broadcast and wireless broadcast with relays.
%\bibliographystyle{ieeetr}
%\bibliography{reference.bib}

\begin{thebibliography}{99}
\bibitem{Yu&Parastoo2018}
M.Yu and P. Sadeghi, ``Approximating throughput and packet decoding delay in linear network coded wireless broadcast,'' \emph{IEEE ITW}, Guangzhou, China, 2018.

\bibitem{Ioannis2017}
I. Chatzigeorgious and A. Tassi, ``Decoding delay performance of random linear network coding for broadcast,'' \emph{IEEE Trans. Vehicular Technology}, vol. 66, no. 8, 2017.

\bibitem{Atilla2013}%3
B. T. Swapna, A. Eryilmaz, and N. B. Shroff, ``Throughput-delay analysis of random linear network coding for wireless broadcasting,'' \emph{IEEE Trans. Inf. Theory}, vol. 59, no. 10, 2013.

\bibitem{Emmanouil2016}%4
E. Skevakis and I. Lambadaris, ``Decoding and file transfer delay balancing in network coding broadcast,'' \emph{IEEE ICC}, Kuala Lumpur, Malaysia, 2016.

\bibitem{NanXie2013}%5
N. Xie and S. Weber, ``Network coding broadcast delay on erasure channels,'' \emph{ITA Workshop}, San Diego, CA, 2013.

\bibitem{SkevakisL2016}%6
E. Skevakis and I. Lambadaris, ``Optimal control for network coding broadcast,'' \emph{IEEE GLOBECOM}, Washington, DC, 2016.

\bibitem{SkevakisL2017}%7
E. Skevakis and I. Lambadaris, ``Delay optimal scheduling for network coding broadcast,'' \emph{IEEE ICC}, Paris, France, 2017.

\bibitem{Lucani2012}%new
D. E. Lucani, M. Medard, M. Stojanovic, ``On coding for delay -- network coding for time-division duplexing,'' \emph{IEEE Trans. Inf. Theory}, vol. 58, no. 4, 2012.

\bibitem{Tsimbalo2018}%new
E. Tsimbalo, A. Tassi, R. J. Piechocki, ``Reliability of multicast under random linear network coding,''. \emph{IEEE Trans. Commu.}, vol. 66, no. 6, 2018.

\bibitem{Sadeghi2010}%8
P. Sadeghi, R. Shams and Danail Traskov, ``An optimal adaptive network coding scheme for minimizing decoding delay in broadcast erasure channels,'' \emph{EURASIP J. Wireless Commun. and Netw.}, 2010.

\bibitem{Aboutorab&Parastoo2014}%9
N. Aboutorab, P. Sadeghi and S. Sorour, ``Enabling a tradeoff between completion time and decoding delay in instantly decodable network coded systems,'' \emph{{IEEE} Trans. Commun.}, vol. 62, no. 4, 2014.

\bibitem{Yu&Parastoo2014}%10
M. Yu, N. Aboutorab and P. Sadeghi, ``From instantly decodable to random linear network coded broadcast,'' \emph{{IEEE} Trans. Commun.}, vol. 62, no. 11, 2014.

\bibitem{Sameh2015}%11
S. Sorour and S. Valaee, ``Completion delay minimization for instantly decodable network codes,'' \emph{{IEEE/ACM} Trans. Netw.}, vol. 23, no. 5, 2015.

\bibitem{Wang2018}%
J. Wang, K. Xu, Y. Xu and D. Zhang, ``Pseudo-systematic decoding of hybrid instantly decodable network code for wireless broadcasting,'' \emph{{IEEE} Wireless  Commun. Letters}, vol. 7, no. 5, 2018.

\bibitem{Giuseppe&Tomaso2015}%12
G. Cocco, T. de Cola and M. Berioli, ``Performance analysis of queueing systems with systematic packet-level coding,'' \emph{{IEEE} ICC}, London, UK, 2015.

\bibitem{Ho2006}%13
T. Ho, M. M{\'{e}}dard, R. Koetter, D. R. Karger, M. Effros, J.Shi and B. Leong, ``A random linear network coding approach to multicast,'' \emph{{IEEE} Trans. Inf. Theory}, vol. 52, no. 10, 2006.

\bibitem{Sameh2010}%14
S. Sorour and S. Valaee, ``On minimizing broadcast completion delay for instantly decodable network coding,'' \emph{IEEE ICC}, Cape Town, South Africa, 2010.

\bibitem{Hou2016}%15
H. Hou, K. W. Shum, M. Chen and H. Li, ``BASIC codes: low-complexity regenerating codes for distributed storage systems,'' \emph{IEEE Trans. Inf. Theory}, vol. 62, no. 6, Jun. 2016.

\bibitem{Sun_TIT19}%16
H. Tang, Q. T. Sun, Z. Li, X. Yang and K. Long, ``Circular-shift linear network coding,'' \emph{IEEE Trans. Inf. Theory}, vol. 65, no. 1, 2019.

\bibitem{Sun_Commu19}%17
Q. T. Sun, H. Tang, Z. Li, X. Yang and K. Long, ``Circular-shift linear network codes with arbitrary odd block lengths,'' \emph{{IEEE} Trans. Commun.}, vol. 67, no. 4, 2019.

\bibitem{Sun_Commu16}%18
Q. T. Sun, X. Yang, K. Long, X. Yin and Z. Li, ``On vector linear solvability of multicast networks,'' \emph{{IEEE} Trans. Commun.}, vol. 64, no. 12,  2016.

\bibitem{Sun_letter19}%19
H. Tang, Q. T. Sun, X. Yang and K. Long, ``On encoding and decoding of circular-shift linear network codes,`` \emph{IEEE Commun. Letters}, vol. 23, no. 5, 2019.

\end{thebibliography}
\begin{appendix}
     \subsection{ Proof of Theorem 1 }
     \label{appendix:proof_thm:GF2_scheme_convergence}
Define two new random variables $N_r$ and $Y_{r,N_r}$ as follows. $N_r$ ($N_r \geq P$) represents the number of \emph{successfully} received packets at receiver $r$ till $r$ can recover all $P$ original packets from them, and $Y_{r,N_r}$ ($Y_{r,N_r} \geq P$) represents the number of packets transmitted by the source till receiver $r$ successfully obtains $N_r$ packets. Equivalently,
\begin{align}
\label{eqn:pmf_RLNC_GF2_2nd_approach}
\mathrm{Pr}(D_r^{\mathrm{GF}(2)} = d)
&= \sum_{n=0}^{d} \mathrm{Pr}(N_r = P+n) \mathrm{Pr}(Y_{r,N_r} = P+d | N_r = P+n) \nonumber \\
&= \sum_{n=0}^{d} \mathrm{Pr}(N_r = P+n) \mathrm{Pr}(Y_{r,P+n} = P+d).
\end{align}%
Note that $D_r^{\mathrm{perf}} = Y_{r, P}$, and generally $Y_{r,P+n}$ also follows the negative binomial distribution and
\begin{equation}
\mathrm{Pr}(Y_{r,P+n} = P + d) = \binom {P+d-1}{P+n-1} p_r^{P+n}(1-p_r)^{d-n}
\end{equation}
     In order to calculate $\mathrm{Pr}(N_r = P+n)$, let $U$ denote the number of uncoded packets among $N_r$ successfully received packets, so that  $\mathrm{Pr}(N_r = P+n)$ can be further calculated by conditioning on $U$. In the case $U = P$, $\mathrm{Pr}(N_r = P | U = P) = 1$ and $\mathrm{Pr}(N_r = P + n | U = P) = 0$ for $n > 0$. In the case $U = u < P$, write
\begin{equation}
\label{eqn:definition_A}
A_{P-u} = \prod_{j = 1}^{P-u}(1 - 2^{-j}) = \prod_{j = 0}^{P-u-1}(1 - 2^{u+j-P}),
\end{equation}
so that
\begin{align}
\label{eqn:conditional_pmf_num_successful_packets_till_P_recoverable}
\mathrm{Pr}(N_r = P + n | U = u) = &
\left\{\begin{matrix}
A_{P-u}, & n = 0 \\
A_{P-u} \sum_{u \leq k_1 \leq k_2 \leq \ldots \leq k_{n} < P} 2^{(k_1-P)+(k_2-P)+\ldots+(k_{n}-P)}, & n > 0
\end{matrix}\right. \nonumber \\
= &\left\{\begin{matrix}
A_{P-u}, & n = 0 \\
A_{P-u} \sum_{0 < k_1 \leq k_2 \leq \ldots \leq k_{n} \leq P-u} 2^{-(k_1+k_2+\ldots+k_{n})}, & n > 0
\end{matrix}\right.
\end{align}
As $U$ follows the binomial distribution with parameter $(P, p_r)$, \begin{align}
\label{eqn:pmf_num_successful_packets_till_P_recoverable}
\mathrm{Pr}(N_r = P + n) = & \sum_{u = 0}^{P} \mathrm{Pr}(U = u) \mathrm{Pr}(N_r = P + n | U = u) \nonumber \\
= & \sum_{u = 0}^{P} \binom{P}{u} p_r^u(1-p_r)^{P-u}\mathrm{Pr}(N_r = P+n | U = u).
\end{align}
We next analyze the expression of $\mathrm{Pr}(N_r = P + n | U = u)$ in (\ref{eqn:conditional_pmf_num_successful_packets_till_P_recoverable}). For $j \geq 1$, write $A_{j1}' = 1$ and recursively define $A_{jk}' = \sum_{j' = 1}^{j} 2^{-j'}A_{j'(k-1)}'$ for $k > 0$. When $n > 0$ and $u < P$, $\mathrm{Pr}(N_r = P + n | U = u)$ can be expressed in terms of $A_{jk}'$ as
\begin{equation}
\label{eqn:conditional_pmf_num_successful_packets_till_P_recoverable_concise}
\mathrm{Pr}(N_r = P + n | U = u) = A_{P-u}\sum_{j' = 1}^{P-u} 2^{-j'}A_{j'n}',
\end{equation}
where $A_{P-u}$ is defined in (\ref{eqn:definition_A}). %
% Because it can be inductively shown that for $j \geq 1$ and $k \geq 2$,
% \[
% A_{jk}' = \sum_{j' = 1}^{j} 2^{-j'}A_{j(k-1)}' < \sum_{j' = 1}^{j} 2^{-j'}A_{j(k-2)}' = A_{j(k-1)}',
% \]
% $\mathrm{Pr}(N_r = P + n | U = u)$ is monotonically decreasing and tends to $0$ when $n$ increases.%
Because $A_{jk}'$ can be alternatively expressed as
\[
A_{jk}' = 2^{-j}A_{j(k-1)}' + A_{(j-1)k}',
\]
it can be inductively deduced that $\mathrm{Pr}(N_r = P + n | U = u)$ exponentially decreases and converges to $0$ when $n$ increases, and $\mathrm{Pr}(N_r = P + n | U = u)$ also converges as $P - u$ increases. For instance, Table \ref{table:conditional_prob_value} lists the probability $\mathrm{Pr}(N_r = P + n | U = u)$ under different choices of $u$ and $n$. As a result, when $P - u$ is larger than a threshold $\delta_u$ (say, e.g., $\delta_u = 20$), we may assume
\begin{equation}
\mathrm{Pr}(N_r = P + n | U = u, P - u \geq \delta_u) = \mathrm{Pr}(N_r = \delta_u + n | U = 0) = \bar{A}_{n},
\end{equation}
where $\bar{A}_{n}$ is equal to $A_{\delta_u}$ when $n = 0$ and denotes $A_{\delta_u}\sum_{j'=1}^{\delta_u} 2^{-j'}A_{j'n}'$ when $n > 0$ for simplicity.

Moreover, because $\sum_{u = P - \delta_u}^P \mathrm{Pr}(U = u) = \sum_{u = P - \delta_u}^{P} \binom{P}{P - \delta_u} p_r^u(1-p_r)^{P-u}$ also tends to $0$ as $P$ increases, we may assume that when $P$ is large enough,
\[
\mathrm{Pr}(N_r = P + n) = \bar{A}_n.
\]
It turns out that when $P$ is large enough, Eq. (\ref{eqn:pmf_RLNC_GF2_2nd_approach}) can be reduced to
\begin{equation}
\label{eqn:pmf_Xr_GF2_concise_form}
\mathrm{Pr}(D_r^{\mathrm{GF}(2)} = d)
= \sum_{n=0}^{d} \bar{A}_n \mathrm{Pr}(Y_{r,P+n} = P+d).
\end{equation}

Similar to (\ref{eqn:expection_D_general_formula}), $\mathbb{E}[D_{\max}^{\mathrm{GF}(2)}] =  \sum_{d \geq 0} \left(1 - \prod_{1 \leq r \leq R} \mathrm{Pr}(D_r^{\mathrm{GF}(2)} \leq d)\right)$, it remains to further analyze $\mathrm{Pr}(D_r^{\mathrm{GF}(2)} \leq d)$. Stemming from (\ref{eqn:pmf_Xr_GF2_concise_form}), we have
\begin{align*}
\mathrm{Pr}(D_r^{\mathrm{GF}(2)} \leq d) = &\sum_{d'=0}^d\sum_{n=0}^{d'} \bar{A}_n\mathrm{Pr}(Y_{r,P+n} = P+d') \\
= &\sum_{n=0}^{d} \bar{A}_n\sum_{d'=n}^d \mathrm{Pr}(Y_{r,P+n} = P+d') \\
= &\sum_{n=0}^{d} \bar{A}_n I_{p_r}(P+n, d-n+1)
\end{align*}
Because $I_{p_r}(P+n, d - n + 1) \leq I_{p_r}(P, d + 1)$ for every $0 \leq n \leq d$, and $\bar{A}_n$ decays exponentially to $0$ as $n$ increases, there exists a bounded integer $\delta_n$ subject to
\begin{equation}
\mathrm{Pr}(D_r^{\mathrm{GF}(2)} \leq d) \geq I_{p_r}(P+\delta_n, d - \delta_n + 1)
\end{equation}
for all receiver $r$ and all $P$ and $d \geq \delta_n$. %
It turns out that
\begin{equation}
\mathbb{E}[D_{\max}^{\mathrm{GF}(2)}] \leq  \delta_n + \sum_{d \geq \delta_n} \left(1 - \prod_{1 \leq r \leq R} I_{p_r}(P+\delta_n, d - \delta_n + 1)\right)
\end{equation}
Recall that $\mathbb{E}[D_{\max}^{\mathrm{perf}}] = \sum_{d \geq 0} \left(1 - \prod_{1 \leq r \leq R} I_{p_r}(P, d + 1)\right)$. Because when $P$ increases, the difference between $I_{p_r}(P, d+1)$ and $I_{p_r}(P+\delta_n, d-\delta_n+1)$ tends to $0$ for all $d \geq \delta_n$, we can now conclude that $E[D_{\max}^{\mathrm{GF}(2)}]/P$ converges to $E[D_{\max}^{\mathrm{perf}}]/P$ when $P$ goes to infinity.

\begin{table}
\center
\caption{The value of $\mathrm{Pr}(N_r = P + n | U = u)$ under different choices of $u$ and $n$}
\label{table:conditional_prob_value}
\begin{tabular}{|l|l|c|c|c|c|c}
  \hline
  \diagbox{$P-u$}{$n$}   & 0    & 1      & 5                      & 10                     & 20  \\
     \hline
  1  & 0.5   & 0.25   & $1.5625\times 10^{-2}$ & $4.8828\times 10^{-4}$ & $4.7684\times 10^{-7}$  \\
  5  & 0.298 & 0.2887 & $2.9395\times 10^{-2}$ & $9.4518\times 10^{-4}$ & $9.2387\times 10^{-7}$\\
  10 & 0.2891 & 0.2888 & $3.0256\times 10^{-2}$ & $9.7466\times 10^{-4}$ & $9.5274\times 10^{-7}$\\
  15 & 0.2888 & 0.2888 & $3.0283\times 10^{-2}$ & $9.7558\times 10^{-4}$ & $9.5364\times 10^{-7}$ \\
  20 & 0.2888 & 0.2888 & $3.0284\times 10^{-2}$ & $9.7561\times 10^{-4}$ & $9.5367\times 10^{-7}$ \\
  \hline
\end{tabular}
\end{table}

\subsection{Proof of Lemma \ref{lemma:no_matrices_circular_shift_RLNC}}
\label{appendix:lemma_proof_matrix_number}
Since $\gamma_{Jj'}$, $1 \leq j' \leq P$, are independently and uniformly chosen from GF($q$), $\mathrm{Pr}(\mathrm{rank}(\mathbf{F}_{J,\gamma}) = J) = 1 - (1/q)^{P-J+1}$. Since $q \leq 1/p_0$,
\begin{equation}
\mathrm{Pr}(\mathrm{rank}(\mathbf{F}_{J,\gamma}) = J) \leq 1 - p_0^{P-J+1}.
\end{equation}

It remains to prove that $\mathrm{Pr}(\mathrm{rank}(\mathbf{F}_{J,\mathbf{\Gamma}}) = JL) \geq 1 - p_0^{P-J+1}$. %
Since $L+1$ is a prime with primitive root $2$, we have the following properties to utilize (See, e.g., \cite{Sun_TIT19})
\begin{itemize}
\item there is an element in $\mathrm{GF}(2^L)$, denoted by $\beta$, with multiplicative order $L+1$, \emph{i.e.}, $\beta^{L+1} = 1$.
\item For a polynomial $f(x)$ over GF($2^L$), if the evaluation $f(\beta) \neq 0$, then $f(\beta^l) \neq 0$ for all $1 \leq \beta \leq L$.
\end{itemize}
Based on the above properties and in the same way to prove Theorem 1 in \cite{Sun_Commu19}, it can be proved that $\mathrm{rank}(\mathbf{F}_{J,\mathbf{\Gamma}}) = JL$ if and only if the $P\times J$ matrix $[\beta_{jj'}]_{1\leq j\leq J, 1 \leq j' \leq P}$ over GF($2^L$) has full rank $J$, where
\begin{equation}
\beta_{jj'} =%
\left\{\begin{array}{ll}
\beta^{l_{jj'}}, & \mathrm{if}~\mathbf{\Gamma}_{jj'} = \mathbf{G}\mathbf{C}_{L+1}^{l_{jj'}}\mathbf{H} \\
0, & \mathrm{if}~\mathbf{\Gamma}_{jj'} = \mathbf{0}
\end{array}\right.
\end{equation}

Consequently, it is equivalent to show that
\begin{equation}
\mathrm{Pr}(\mathrm{rank}(\mathbf{F}_{J,\mathbf{\beta}}) = J) \geq 1 - p_0^{P-J+1},
\end{equation}
where $\mathbf{F}_{J,\mathbf{\beta}} = [\beta_{jj'}]_{1\leq j \leq J,1\leq j' \leq P}$ is a $P\times J$ matrix defined over GF($2^L$) with $\mathbf{\beta}_{jj'} \in \{0, \beta, \beta^2, \ldots, \beta^{L+1}\} \subseteq \mathrm{GF}(2^{L'})$ subject to
\begin{itemize}
\item the $P\times (J-1)$ submatrix $[\beta_{jj'}]_{1\leq j < J,1\leq j' \leq P}$ is full rank,
\item every $\beta_{Jj'}$ is randomly selected from $\{0, \beta, \beta^2, \ldots, \beta^{L+1}\}$ according to the distribution
    \begin{align}
    \mathrm{Pr}(\beta_{Jj'} = \beta') =
    &\left\{\begin{matrix}
    p_0, & \beta' = 0 \\
    \frac{1-p_0}{L+1}, & \beta' = \beta^l, 1\leq l \leq L+1
    \end{matrix}\right.
    \end{align}
\end{itemize}

As $p_0$ is a rational number, write $p_0 = a/b$. Expand the set $\{0, \beta, \beta^2, \ldots, \beta^{L+1}\}$ into a \emph{multiset} $\mathcal{M}_\beta$ of cardinality $b(L+1)$ as follows:
\begin{itemize}
\item the multiplicity of $0$ is $a(L+1)$, \emph{i.e.}, $\mathcal{M}_\beta$ contains $a(L+1)$ duplicate $0$.
\item for $1 \leq l \leq L+1$, the multiplicity of $\beta^l$ is $b-a$, \emph{i.e.}, $\mathcal{M}_\beta$ contains $b-a$ duplicate $\beta^l$.
\end{itemize}
In this way,
\begin{equation}
\mathrm{Pr}(\mathrm{rank}(\mathbf{F}_{J,\beta}) = J) = 1 - \frac{A_\beta}{(b(L+1))^P},
\end{equation}
where $A_\beta$ represents the number of (possibly duplicate) column vectors in $\{[\beta_{Jj'}]_{1\leq j'\leq P}: \beta_{Jj'} \in \mathcal{M}_\beta \}$ that can be written as a GF($2^L$)-linear combination of the $J-1$ column vectors $[\beta_{1j'}]_{1\leq j'\leq P}, \ldots, [\beta_{(J-1)j'}]_{1\leq j'\leq P}$.

As the $P \times (J-1)$ matrix $[\beta_{jj'}]_{1\leq j < J, 1 \leq j' \leq P}$ has rank $J-1$, it contains $J-1$ linearly independent rows. Without loss of generality, assume $\mathrm{rank}([\beta_{jj'}]_{1\leq j,j' < J}) = J-1$. Thus, for each of the $(b(L+1))^{J-1}$ column vectors $[\beta_{Jj'}]_{1\leq j' < J}$ with $\beta_{Jj'} \in \mathcal{M}_\beta$, there are unique $\alpha_1, \alpha_2, \ldots, \alpha_{J-1} \in \mathrm{GF}(2^L)$ subject to $\sum_{1\leq j < J} \alpha_j[\beta_{jj'}]_{1\leq j' < J} = [\beta_{Jj'}]_{1\leq j' < J}$. %
Moreover, for $J \leq j' \leq P$, $\sum_{1\leq j < J} \alpha_j\beta_{jj'}$ may or may not be in $\mathcal{M}_\beta$. If $\sum_{1\leq j < J} \alpha_j\beta_{jj'} \in \mathcal{M}_\beta$, then it has at most $a(L+1)$ duplicates in $\mathcal{M}_\beta$, since we assume $p_0 = \frac{a}{b} \geq \frac{1}{L+2}$ so that $a(L+1) \geq b-a$. Thus, $A_\beta$ is upper bounded by $(b(L+1))^{J-1}(a(L+1))^{P-J+1}$ and hence,
\begin{equation}
\mathrm{Pr}(\mathrm{rank}(\mathbf{F}_{J,\beta}) = J) = 1 - \frac{A_\beta}{(b(L+1))^P} \geq 1 - (a/b)^{P-J+1}.
\end{equation}

\subsection{Proof of Theorem \ref{thm:completion_delay_circular-shift_RLNC}}
\label{appendix:thm_proof_completion_delay_circular-shift_RLNC}
It remains to prove $\mathrm{Pr}(D_r^{\mathrm{circ}} \leq d) \geq \mathrm{Pr}(D_r^{\mathrm{GF}(q)} \leq d)$ for all $d \geq 0$. %
Same as the discussion in Sec.II-B, denote by $U_{r}$ the number of successfully received uncoded packets at receiver $r$ in phase one, and assume that $U_r = u$ with $u < P$. Since when $u + d < P$, $\mathrm{Pr}(D_r^{\mathrm{circ}} \leq d) = \mathrm{Pr}(D_r^{\mathrm{GF}(q)} \leq d) = 0$, we further assume that $d \geq P - u$. Similar to (\ref{eqn:D_r_expression}), $D_r^{\mathrm{circ}}$ and $D_r^{\mathrm{GF}(q)}$ can be respectively expressed as
\begin{equation}
\begin{split}
D_r^{\mathrm{circ}} &= A_1^{\mathrm{circ}} + A_2^{\mathrm{circ}} + \ldots + A_{P-u}^{\mathrm{circ}}, \\
D_r^{\mathrm{GF}(q)} &= A_1^{\mathrm{GF}(q)} + A_2^{\mathrm{GF}(q)} + \ldots + A_{P-u}^{\mathrm{GF}(q)},
\end{split}
\end{equation}
where $A_j^{\mathrm{circ}}$ and $A_j^{\mathrm{GF}(q)}$ respectively represent the number of coded packets obtained by receiver $r$, under the proposed circular-shift RLNC scheme and under the conventional RLNC scheme over GF($q$), during the process that the number of innovative packets at receiver $r$ increases from $u+j-1$ to $u+j$. Thus, $A_j^{\mathrm{circ}}$ and $A_j^{\mathrm{GF}(q)}$ follow the geometric distribution with the respective parameter ${p'}_{r,u+j-1}^{\mathrm{circ}} = p_r\mathrm{Pr}\left(\mathrm{rank}(\mathbf{F}_{u+j,\mathbf{\Gamma}}) = (u+j)L\right)$ and ${p'}_{r,u+j-1}^{\mathrm{GF}(q)} = p_r\mathrm{Pr}\left(\mathrm{rank}(\mathbf{F}_{u+j,\gamma}) = u+j \right)$, where $\mathbf{F}_{u+j,\mathbf{\Gamma}}$ and $\mathbf{F}_{u+j,\gamma}$ are in line with the definition above Lemma \ref{lemma:no_matrices_circular_shift_RLNC} with the setting $J = u+j$. %
Consequently,
\begin{equation}
\begin{split}
\mathrm{Pr}(D_r^{\mathrm{circ}} = d | U_r = u) &=
\sum_{\mathbf{a} \in \mathcal{A}_{P-u,d}} \prod_{j=1}^{P-u}(1 - {p'}_{r, u+j-1}^{\mathrm{circ}})^{a_j-1}{p'}_{r, u+j-1}^{\mathrm{circ}}, \\
\mathrm{Pr}(D_r^{\mathrm{GF}(q)} = d | U_r = u) &=
\sum_{\mathbf{a} \in \mathcal{A}_{P-u,d}} \prod_{j=1}^{P-u}(1 - {p'}_{r, u+j-1}^{\mathrm{GF}(q)})^{a_j-1}{p'}_{r, u+j-1}^{\mathrm{GF}(q)},
\end{split}
\end{equation}
where $\mathcal{A}_{P-u,d}$ is defined in the same way as in (\ref{eqn:conditional_pmf_RLNC_GF2}), that is, every $(P-u)$-tuple $\mathbf{a}=(a_{1},\ldots, a_{P-u})$ of positive integers satisfies $a_{1} + \ldots + a_{P-u} = d$. %

For $0 \leq n < P$ and $d \geq n$, define a function $f_{n,d}$ of an $n$-tuple $\mathcal{X}_n = (x_1, \ldots, x_n)$ of variates according to
\begin{equation}
f_{n,d}(\mathcal{X}_n) = \sum_{\mathbf{a} \in \mathcal{A}_{n,d}} \prod_{j=1}^n(1 - x_j)^{a_j-1}x_j.
\end{equation}
Thus,
\begin{equation}
\begin{split}
\mathrm{Pr}(D_r^{\mathrm{circ}} \leq d | U_r = u) &= \sum_{d' = P-u}^d f_{P-u,d'}(\wp^\mathrm{circ}), \\
\mathrm{Pr}(D_r^{\mathrm{GF}(q)} \leq d | U_r = u) &= \sum_{d'=P-u}^d f_{P-u,d'}(\wp^{\mathrm{GF}(q)}),
\end{split}
\end{equation}
where $\wp^\mathrm{circ} = ({p'}_{r, u}^{\mathrm{circ}}, {p'}_{r, u+1}^{\mathrm{circ}},\ldots, {p'}_{r, P-1}^{\mathrm{circ}})$, and $\wp^{\mathrm{GF}(q)} = ({p'}_{r, u}^{\mathrm{GF}(q)}, {p'}_{r, u+1}^{\mathrm{GF}(q)},\ldots, {p'}_{r, P-1}^{\mathrm{GF}(q)})$ %

For $d \geq n = 1$, $\sum_{d' = 1}^d f_{1,d'}(x) = \sum_{d' = 1}^d (1-x)^{d'-1} x = 1 - (1-x)^d$. Thus, for $x < 1$,
\begin{equation}
\label{eqn:derivative_single_variate}
\left(\sum_{d' = 1}^d f_{1,d'}(x)\right)' = d(1-x)^{d-1} > 0,
\end{equation}
where the symbol $'$ means the derivative operation. %
For $d \geq n \geq 2$ and arbitrary $1 \leq j \leq n$, by conditioning on the possible value $d''$ for $a_j$, one may readily deduce
\begin{align}
& \sum_{d'=n}^d f_{n,d'}(\mathcal{X}_{n}) \nonumber \\
= &\sum_{d'=n}^d\sum_{d'' = 1}^{d'+1-n} f_{n-1,d'-d''}(\mathcal{X}_{n}\backslash x_j)f_{1,d''}(x_j) \nonumber \\
= &\sum_{d'=n}^d f_{n-1,d'-1}(\mathcal{X}_{n}\backslash x_j) \sum_{d'' = 1}^{d+1-d'} f_{1,d''}(x_j),
\end{align}
where the last equation can be obtained by changing the way to enumerate all $\frac{(d+2-n)(d+1-n)}{2}$ possible choices. %
Thus, for $0 < x_1, \ldots, x_n < 1$,
\begin{equation}
\frac{\partial\left(\sum_{d' = n}^d f_{n,d'}(\mathcal{X}_n)\right)}{\partial{x_j}} = \frac{\sum_{d'=n}^d f_{n-1,d'-1}(\mathcal{X}_{n}\backslash x_j)\partial\left(\sum_{d'' = 1}^{d+1-d'} f_{1,d''}(x_j)\right)}{\partial{x_j}} > 0,
\end{equation}
where the last inequality stems from (\ref{eqn:derivative_single_variate}) and the fact that $f_{n,d}(\mathcal{X}_n) > 0$ for all $d \geq n \geq 1$ and $0 < x_1, \ldots, x_n < 1$. %
It turns out that for $\mathcal{X}_n = (x_1, \ldots, x_n)$, $\mathcal{X}_n' = (x_1', \ldots, x_n')$ with $0 < x_j \leq x_j' < 1$ for all $1 \leq j \leq n$, $\sum_{d' = n}^d f_{n,d'}(\mathcal{X}_n) \leq \sum_{d' = n}^d f_{n,d'}(\mathcal{X}_n')$.

Since ${p'}_{r,u+j-1}^{\mathrm{circ}} \geq {p'}_{r,u+j-1}^{\mathrm{GF}(q)}$ for all $1 \leq j \leq P-u$,
\begin{equation}
\begin{split}
\mathrm{Pr}(D_r^{\mathrm{circ}} \leq d | U_r = u) = &\sum_{d' = P-u}^d f_{P-u,d'}(\wp^\mathrm{circ}) \\
\geq &\sum_{d'=P-u}^d f_{P-u,d'}(\wp^{\mathrm{GF}(q)}) = \mathrm{Pr}(D_r^{\mathrm{GF}(q)} \leq d | U_r = u),
\end{split}
\end{equation}
and hence
\begin{equation}
\begin{split}
\mathrm{Pr}(D_r^{\mathrm{circ}} \leq d)  = &\sum_{u = \max\{0, P-d\}}^{P-1} \mathrm{Pr}(D_r^{\mathrm{circ}} \leq d | U_r = u)\mathrm{Pr}(U_r = u) \\
\geq & \sum_{u = \max\{0, P-d\}}^{P-1}\mathrm{Pr}(D_r^{\mathrm{GF}(q)} \leq d | U_r = u)\mathrm{Pr}(U_r = u) = \mathrm{Pr}(D_r^{\mathrm{GF}(q)} \leq d).
\end{split}
\end{equation}

\subsection{List of Notation}

\label{List_of_Notation}
\begin{xtabular}{p{16mm}l}
$P$:     &     the number of original packets.  \\
$R$:     &     the number of receivers.    \\
$M$:   &      the number of symbols in a packet. \\
$L$:    &   the number of bits representing a symbol. \\
$p_r$:       &   the probability of successfully receiving a packet at receiver $r$. \\
$D$:   &   the completion delay of an RLNC scheme.  \\
$D_r$:   &   the completion delay of an RLNC scheme corresponding to receiver $r$.  \\
$\mathbf{m}_{j}$:  &   the transmitted packet, which is a row vector of $M/L$ symbols. \\
$\mathbf{m}_{j}'$:  &  the received packet. \\
$\gamma$:   &   the coding coefficient selected from GF($2^L$) in the conventional RLNC scheme.\\
$\mathbf{\Gamma}$:   &   the $L\times L$ coding coefficient matrix over GF($2$) in the proposed RLNC scheme.\\
$U_r$:  &   the number of successfully received uncoded packets by receiver $r$ in phase I.\\
$\mathbf{I}_{L}$: & the $L\times L$ identity matrix. \\
$\mathbf{C}_{L+1}$: & the $(L+1)\times (L+1)$ cyclic permutation matrix. \\
$\mathbf{G}$: & the $L\times (L+1)$ matrix defined in (\ref{eq:definition_GH}).\\
$\mathbf{H}$: & the $(L+1)\times L$ matrix defined in (\ref{eq:definition_GH}).\\
$\mathcal{C}$: &  the set of coding coefficients defined in (\ref{eq:definition_C_cal}).\\
$p_0$:  &   the probability of a random coding coefficient to be $\mathbf{0}$ in the proposed scheme.\\
$\circ$: & the symbol-wise multiplication defined in (\ref{eqn:circ-symbol-definition}).
\end{xtabular}

\end{appendix}

\section{Acknowledgement}
The authors would like to appreciate the valuable suggestions by the associate editor as well as anonymous reviewers to help improve the quality of the paper.

\end{document}